\documentclass[doublespace,
ResearchPaper,harvardrefstyle,onecolumn]{CSIRO_LaTeX_Template}

\usepackage{subcaption}

\begin{document}

\journaltype{Research Paper}

\jwebadd{https://doi.org/10.1071/xxxx}

\journalname{Journal name}

\title{Leveraging MTG-FCI fire observations for event-based fire behavior monitoring from near-real-time operation to seasonal analysis}

\versorh{Paugam et al}
\rectorh{www.publish.csiro.au/xx}

\author[1,*\orc{0000-0001-6429-6910}]{Ronan Paugam}
\author[2\orc{0000-0002-6244-0648}]{Jean-Baptiste Filippi}
\author[3\orc{0000-0002-4325-3804}]{Akli Benali}
\author[4\orc{0009-0009-6877-5084}]{Jorge Gomes}
\author[5\orc{0000-0001-9685-3742}]{Weidong Xu}
\author[6\orc{0000-0002-0643-2643}]{Emanuel Dutra}
\author[7]{Francois Andre}
\author[7\orc{0000-0001-6935-1106}]{Damien Boulanger}
\author[7]{Vianney Retornard}
\author[8\orc{0000-0002-7690-5346}]{Andrea Meraner}
\author[9\orc{0009-0004-7463-0176}]{Julia Harvie}
\author[10\orc{0000-0001-9142-9766}]{Victor Penot}
\author[11\orc{0000-0001-5559-5732}]{Cyrielle Denjean}

\affil[1]{CERTEC, Department of Chemical Engineering, Universitat Politècnica de Catalunya Barcelona Tech, Campus Diagonal Besòs Edifici I, Eduard Maristany, 16, Barcelona, Spain}{}
\affil[2]{SPE--Sciences Pour l’Environnement, Université de Corse, CNRS, Corte, France}{}
\affil[3]{Centro de Estudos Florestais, Instituto Superior de Agronomia, Universidade de Lisboa, Portugal}{}
\affil[4]{Virtual Operations Support Team Portugal; Virtual Operations Support Team Europe}{}
\affil[5]{King's College London, Leverhulme Centre for Wildfires, Environment and Society, UK}{}
\affil[6]{Instituto Português do Mar e da Atmosfera, Lisbon, Portugal}{}
\affil[7]{OMP-ATLAS, CNRS, France}{}
\affil[8]{EUMETSAT, Darmstadt, Germany}{}
\affil[9]{Great Lakes Forestry Centre, Canadian Forest Service, Natural Resources Canada, Sault Ste. Marie, Ontario, Canada}{}
\affil[10]{URFM, INRAE, Avignon, France}{}
\affil[11]{Météo-France, CNRS, Univ. Toulouse, CNRM, Toulouse, France}{}

\corau{*Corresponding author. Email: ronan.paugam@upc.edu}

%\corautwo{$^{\#}$These authors contributed equally to this paper [If relevant]}

\begin{abstract}
Wildfire monitoring and suppression require timely information on fire behavior, including fire energy release and rate of spread, to support operational decision-making and resource allocation. Active fire products from the Flexible Combined Imager (FCI) aboard the geostationary Meteosat Third Generation (MTG) satellites provide 10-min observations over Europe and Africa. Deriving fire behavior information from these observations requires associating individual hotspot detections into coherent fire events.

We present a Fire Event Tracker (FET) algorithm that performs spatio-temporal clustering of hotspot detections from the LSA-SAF FCI active fire product. The algorithm assigns persistent identifiers to fire events and updates their geometry, fire radiative power, and rate of spread at each 10-min interval. The same parameterization is used for both near-real-time and retrospective processing.
FET was applied retrospectively to the Mediterranean FCI hotspot archive of  2025 and operationally in two near-real-time contexts: wildfire monitoring in Portugal and support of the 2025 SILEX airborne campaign within the EUBURN project, where beside fire monitoring FET products were also used to initialize coupled FOREFIRE–MesoNH simulations for plume forecasting.

Results show that event-based clustering of FCI active fire detections provides a consistent description of fire evolution, enabling both tactical wildfire management and high-frequency seasonal fire analyses.
\end{abstract}

\begin{keywords}
Wildfire, Satellite, Fire Radiative Power, Rate of Spread.
\end{keywords}

\maketitle

\begin{onlinesummarytext}[Online Summary Text :]
Wildfires are becoming more frequent and intense across Europe, creating growing risks for people, infrastructure, and ecosystems. This study introduces a new way to track fires using continuous satellite observations, allowing individual fires to be followed in near real time and revealing how their intensity and spread evolve over time. These advances pave the way to faster, more informed decision-making during emergencies and improved understanding of fire activity over entire seasons.
\end{onlinesummarytext}

\section{Introduction}
%------------------------
Wildfires are increasingly recognized as one of the most disruptive natural hazards in Europe and globally \citep{GAR2022}. Climate projections indicate that changes in precipitation regimes and rising temperatures will substantially alter fire weather conditions. In Europe, recent modeling studies using Coupled Model Intercomparison Project Phase 6 (CMIP6) scenarios suggest that, under moderate forcing, the probability of extreme fire weather may increase by an order of magnitude in any given year compared to the recent past \citep{elGarroussi2024}. These shifts are expected to amplify the frequency and severity of large fire episodes, challenging the preparedness and response capacity of national and regional firefighting agencies.

The operational consequences of these trends are significant. Fire suppression resources --already strained by compound and prolonged events in North America, Australia, and increasingly in Europe-- will face heightened demand for situational awareness and rapid decision-making. Conventional fire monitoring approaches, which primarily rely on ground-based observations and temporally delayed satellite products from sensors such as the Moderate Resolution Imaging Spectroradiometer (MODIS) and the Visible Infrared Imaging Radiometer Suite (VIIRS) operated onboard polar-orbiting satellites, remain challenged in meeting the increasing demand for fire monitoring at high temporal (sub-hourly) and fine spatial (sub-kilometer) resolutions required for current and future fire management operations. This gap is driving the development of near-real-time (NRT) observation systems, nowcasting frameworks, and short-range forecasting chains that integrate multi-sensor fire detection with physics-based models of fire spread and smoke dispersion \citep{valabre_panoptes_2025}. Such systems aim to provide actionable intelligence on fire behavior at the scales required by incident commanders and civil protection authorities.

Over the last decade, research prototypes for real-time fire modeling have matured into pre-operational tools in several fire-prone regions \citep{Kochanski2016}. Examples include coupled atmosphere–fire models capable of simulating fire growth and plume rise \citep{mandel2014,Filippi2025}, as well as satellite-based active fire products that provide detections within minutes to hours of overpass \citep{wooster2012a,wooster2015,Xu2026}. In North America and Australia, these capabilities have begun to transition into operational use, supporting resource allocation, evacuation planning, and safety protocols. In Europe, however, distinct operational constraints exist: wildfires are often shorter-lived than in other fire environments, and occur in densely populated landscapes and in proximity to critical infrastructure. This increases both the urgency of response and the value of timely, high-resolution fire intelligence.

A key aspect of this transition is the improvement of observation systems to support fire behavior monitoring. Advances in satellite capabilities now enable multi-scale observation: polar-orbiting platforms (e.g., VIIRS, Sentinel-3 SLSTR) deliver detailed active fire and burned area products at a spatial resolution up to $400~m$, while geostationary missions (e.g., Meteosat Second Generation MSG-SEVIRI, Himawari-AHI, GOES-R-ABI, and the new Meteosat Third Generation MTG-FCI) provide high-temporal-resolution active fire detection and monitoring suitable for NRT applications. These datasets—particularly FCI observations, which provide a spatial resolution of approximately 1 × 1.5 km over southern Europe --three times finer than its predecessor SEVIRI-- open unprecedented opportunities for monitoring fire-front dynamics, quantifying fire spread rates, and anticipating smoke transport at temporal cadences suitable for operational decision support.
Integrating these multi-sensor observations with physics-based fire spread and coupled fire–atmosphere models represents the next step towards predictive fire intelligence. Such systems have the potential not only to inform suppression strategies and resource allocation in real time, but also to anticipate extreme fire behavior, identify windows of opportunity for control, and mitigate risks to responders and communities. In Europe, where fire events are increasingly affecting urban environment, transport corridors, and energy infrastructure, the demand for such integrated capabilities is particularly acute.
Overall, the combined increase in climate-driven fire risk, advances in remote sensing technologies, and operational demand for rapid decision-making highlights the need to transition toward fully operational NRT fire monitoring and forecasting systems to better protect populations, infrastructure, and ecosystems.
%
%Overall, the convergence of climate-driven fire risk escalation, technological advances in remote sensing, and operational needs for rapid decision-making underscores the urgency of advancing from research prototypes to fully operational NRT fire monitoring and forecasting chains. Strengthening these capabilities will be critical for safeguarding populations, infrastructure, and ecosystems in a fire-prone 21st century.
Recent advances in fire monitoring algorithm have enabled the tracking of fire events using polar‑orbiting hotspot observations from VIIRS, as well as the derivation of associated event‑level metrics such as the Rate of Spread (ROS) \cite{chen2022}.
In this study, we present a step toward the exploitation of geostationary active fire observations for systematic fire event analysis using FCI sensor onboard MTG-I1/Meteosat-12. Its spatial resolution combined with high temporal sampling and low latency provide the opportunity to observe fire activity continuously at a cadence not previously available over Europe. Building on the Fire Radiative Power (FRP) hotspot product delivered by LSA-SAF, we introduce a Fire-Event-Tracker (FET) algorithm that associates individual hotspot detections into persistent fire events using spatio-temporal clustering, and derives event-level time series of fire behavior metrics at a 10-minute temporal resolution.

The manuscript is structured as follows. First, we describe the FCI active fire dataset and the tracking algorithm used to construct fire event histories and associated metrics. We then illustrate three applications of the same algorithm under different data availability contexts. The first is a retrospective analysis of the 2025 fire season over the Mediterranean region, based on the LSA-SAF FCI hotspot archive, demonstrating the ability of FET to generate year-long, high-frequency event-level characterization at the spatial scale of fire regimes. The two other applications highlight NRT deployments during the 2025 fire season. One focuses on operational monitoring in Portugal, conducted in collaboration with the National Authority for Emergency and Civil Protection (ANEPC). 
%where FET was found particularly useful to support the identification and monitoring of fire re-activations. 
The second application concerns the 2025 SILEX airborne campaign, during which NRT fire event detections were used to trigger coupled FOREFIRE–MesoNH simulations to support plume dispersion assessment and flight planning.

%Together, these applications demonstrate how a single, invariant fire event tracking approach applied to geostationary active fire detections can support both retrospective seasonal analyses and NRT operational uses, solely as a function of data completeness, without changes to the underlying processing chain.

\section{Material and Methods}
%------------------------
\subsection{The algorithm}
%~~~~~~

Following development in fire monitoring from Chen et al\citep{chen2022}, FET was developed to ingest and track hotspot detections from the LSA-SAF MTG product \cite{Xu2026}.
FET algorithm inputs are the hotspot product (point vector file containing FRP point values) from the LSA-SAF distribution service. They are available with a latency of approximately 20 minutes and delivered with the 10-min cadence of the sensor. The hotspot product covers the full Earth disc field-of-view of FCI.

Active fire hotspot products may include non-fire detections that must be filtered prior to analysis, as some thermal anomalies arise from sources such as sunglint, gas flaring, or other persistent industrial heat sources rather than biomass burning \cite{giglio2003_MODISFireDestection}. According to \cite{Xu2026}, the FCI hotspot product is delivered with a $88\%$ true positive rate that compares to other similar hotpot products. However, it is recommended to complement it with a temporal persistence filter \cite{Koltunov2012} to reduce commission errors.
A mask of persistent hotspots was generated from the $375~$m VIIRS active fire product \cite{Schroeder2014a}, whose higher spatial resolution improves the localization and temporal consistency of detections, facilitating the identification and exclusion of industrial heat sources.
This was carried out using the 2024 VIIRS archive to reduce the real-time processing load in the 2025 operational application.
The fixed hotspot mask is produced from a density kernel of contiguous fixed clusters, which are defined as having a total lifetime longer than 6 months and a maximum inactivity gap of less than 2 months. The binary fixed hotspot mask is subsequently created by applying a threshold to the cluster density raster (e.g., maximum density $ \geq 12$).

The FET framework is a Python-based, near–real-time system designed to cluster active fire detections into spatio-temporal “events”. The core algorithm applies DBSCAN clustering to the hotspot detections (defined at the center of the flagged pixels) \citep[e.g.][]{su2023} and subsequently uses the alpha shape algorithm to derive a concave polygon around each cluster \citep{chen2022} delimitating the area of interest of the event at a given time to which is assign an FRP equal to the sum of the current active hotspot. We named this polygon the "area of interest" (AOI) and not "fire perimeter" as the polygon resolution inherited from FCI spatial resolution is too coarse to compare to fire perimeter product used during operational fire attack. The final AOI is compared to EFFIS Burnt Area (BA) product in the discussion section of the manuscript.
The DBSCAN clustering is run with a spatial buffer of 2 km, while the concave hull is constructed with an alpha parameter set as the mean distance of the nearest distance between all hotspot that are included in the cluster. This results in an alpha shape parameter that is very close of the 1 km used by \cite{chen2022} when using VIIRS hotspots clusters.

The strategy of the hotspot clustering and event tracking is based on a pool of active hotspot that is updated at every new time step. The lifetime of each hotspot is controlled through the event activity (when event are terminated, hotspots are removed from the pool). At each time step the pool of hotspot is clustered (DBSCAN) and polygons are set (alpha shape) as mentioned above. Each new polygon is then compared to the last polygon of all the active event from the last iteration. The newly set polygon can create a new event, be added to an existing event, or triggers the merging of two active events.
An event termination occurs when no hotspot update happens in the last 48 hours. When an event terminates, the according hotspots are removed from the pool of active hotspots.
At the end of each step of the time integration, a Forward Rate of Spread (FROS) \cite{Benali2023} calculation (See Appendix for details) is performed for each new perimeter in the event with more than 2 perimeters.

From a technical perspective, the alpha shape algorithm is applied only to clusters that contain four points or more. Clusters with exactly three points are directly modeled as a simple polygon. For clusters with two or fewer points, events are represented by a point or a line. Throughout this manuscript, the term “polygon” is used when referring to the geometry associated with an event. However, note that the implementation governing event interactions and evolution is designed to also handle point and line events when they occur.
The sensitivity of the parameters listed above for DBSCAN and the alpha shape algorithm were tuned heuristically by trial and error for operational deployment during the 2025 fire season. A dedicated parametric study is planned for future work.

Resulting events (i) are tracked using unique persistent identifiers, (ii) are updated at every 10-minute time step of the FCI sensor, and (iii) retain the complete history of geometries and attributes from all prior event states. Each event at a given time is written to a single file, which enables simple postprocessing to derive AOIs/polygons and time series of FRP and FROS. Full event states over the domain of interest can be archived at a temporal resolution that is decoupled from the model integration time step, and FET is configured to restart from any stored event state.
In addition to the output dumping interval, the duration of the integration period is configurable and can be tailored to the specific application. 

\begin{figure}
    \centering
    \includegraphics[width=0.7\linewidth]{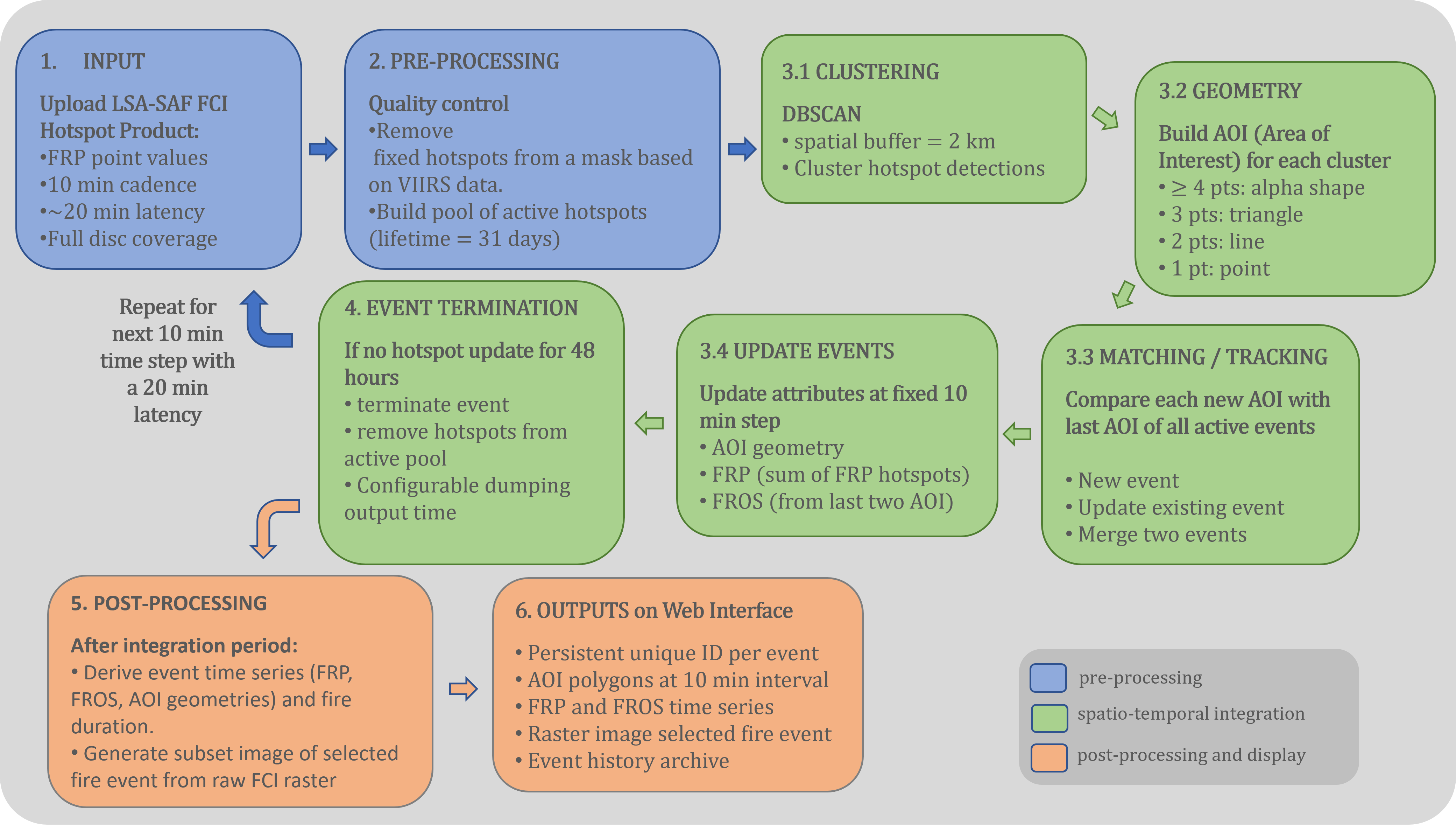}
    \caption{Flow chart of the Fire Event Tracker (FET) algorithm}
    \label{fig:flowChart}
\end{figure}

A schematic representation of the algorithm is provided in Fig. \ref{fig:flowChart}. In addition to the time-integration procedure, a postprocessing stage is included in which each event is associated with supplementary fire behavior metrics, such as the Fire Radiative Energy (FRE)—defined as the temporal integral of the Fire Radiative Power (FRP)—and the fire duration, defined as the time interval between the initiation and the termination of the event.

Figure \ref{fig:flowChart} presents a flowchart that provides a concise overview of the procedural steps of the FET algorithm.

\subsection{The study areas}
%~~~~~~

Three computational domains with distinct spatial extents are considered in this study. As illustrated in Figure \ref{fig:FETConfig}, these domains span a range of spatial scales, from the national scale to scale representative of fire-regime like the Mediterranean basin:
\begin{figure}[ht]
    \centering
    \includegraphics[width=.7\linewidth]{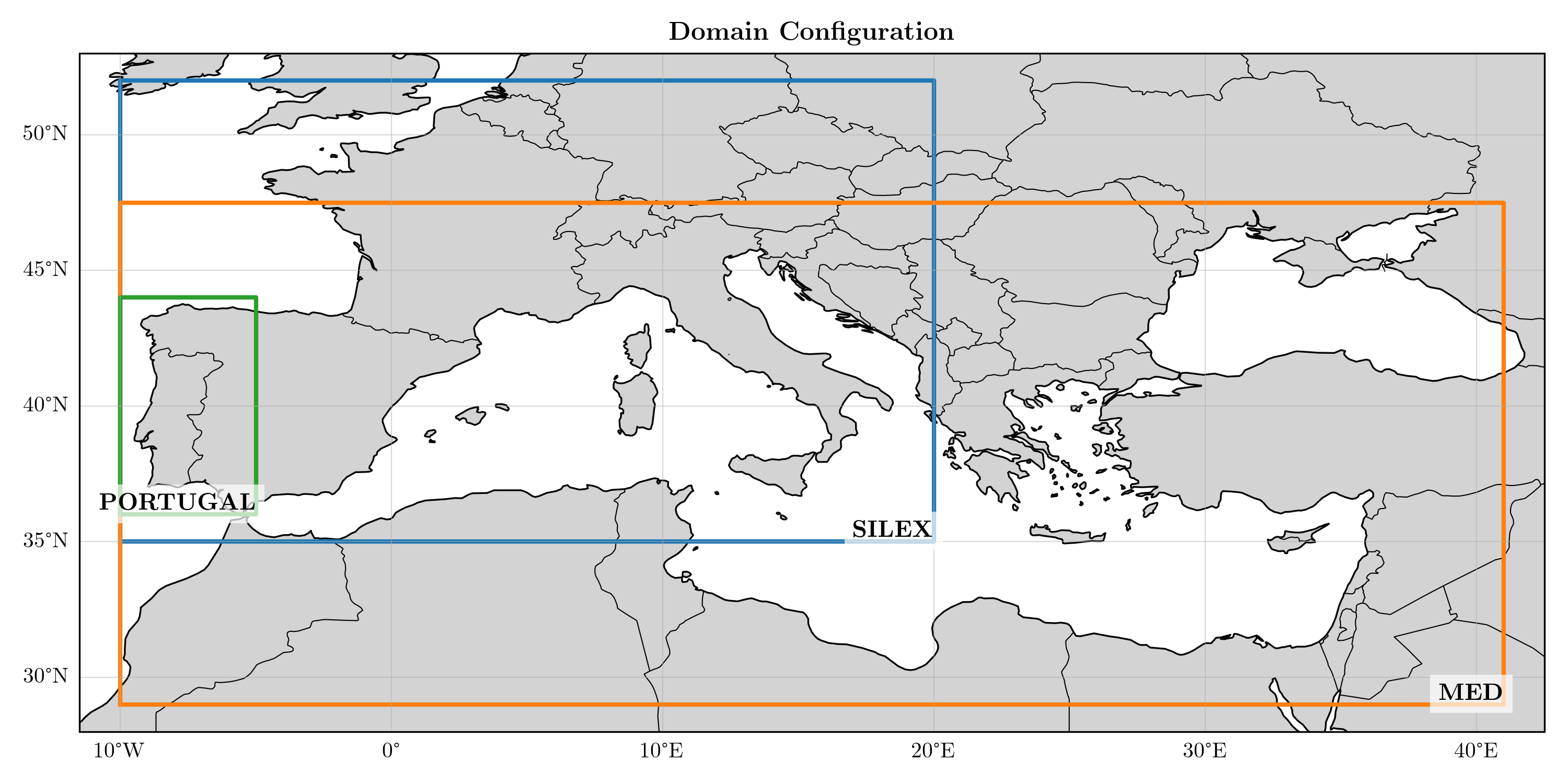}
    \caption{Domain extension of the three configuration set-up of the Fire Event Tracker Algorithm (MED,PORTUGAL and SILEX) presented in this work.}
    \label{fig:FETConfig}
\end{figure}
\begin{itemize}
\item PRT [small]: a NRT operational setup over Portugal to provide fire activity observation to ANEPC to help decision planning,
\item SILEX [medium]: a further NRT operational setup deployed during the Smoke from wILdfires EXperiment (SILEX) airborne campaign where FET was coupled with the FOREFIRE-MESONH fire-atmosphere coupled system to provide fire activity monitoring and plume direction prediction to help the flight authorization delivery.
\item MED [large]: a retrospective analysis of the year 2025 over the Mediterranean basin,
\end{itemize}

The SILEX campaign \cite{Denjean2025} was conducted in summer 2025 as part of the broader EUBURN programme, with the aim of improving the understanding of wildfire emissions, plume development, and their impacts on air quality and climate. The campaign was coordinated by Météo-France/CNRM in collaboration with several European partners, and combined airborne, ground-based, and satellite observations. The SAFIRE ATR-42 research aircraft, equipped with in-situ and remote sensing instruments, carried out 14 scientific flights between mid-July and early August 2025 in southern France regions, targeting both active fire fronts and downwind plumes for in-situ sampling. Measurements encompassed fire behavior monitoring, gas and aerosol properties, and plume rise, providing unprecedented datasets on the composition and evolution of fire plumes in southern Europe.
%Beyond its scientific objectives, SILEX also served to test novel instrumentation, strengthen cross-border collaboration, and support the calibration and validation of satellite fire and atmospheric products. The campaign highlighted both the potential and the logistical challenges of rapid-response airborne operations in complex wildfire environments. The collected datasets will underpin future improvements to emission inventories, air quality forecasting, and coupled fire–atmosphere modeling, and will serve as a foundation for a more extensive follow-up campaign planned under EUBURN in 2027.

\subsection{Data Visualization }
%~~~~~~
Web interfaces were developed to visualize the resulting dataset of spatio-temporal varying AOI associated with fire behavior metrics (see supplementary material).

\section{Results}
%------------------------
This section present three deployments of the FET algorithm in the configuration listed above.

\subsection{Application to a retrospective analysis of the year 2025 - [MED]}
\label{sec:retro2025}
%~~~~~~
A retrospective analysis of the year 2025 was performed using the LSA-SAF FCI hotspot active fire product archive. The FET algorithm was executed sequentially using daily integration time period and a $30~$min dumping interval. This means that fire events for the whole domaine are maintained in memory during each 24-hour integration and every $30~$min the state of all the active events are saved. At the start of the following day, the last events state is loaded and the integration resumed. Every $30~$min time interval, individual event state of each event monitored at 10 min time interval are saved in separate files.

A total of 139,604 individual fire events were detected, resulting in 136,207 distinct final fire events. The difference between the two numbers ($<3\%$) arises from the progressive aggregation and merging of fire perimeters throughout their spatiotemporal evolution.
Out of 136,207 detected events, 29,485 include more than one observation (duration > 10 min) and are more likely to be representative of wildfire incidents. This annual count is lower than the ~50,000 fire events reported for the Mediterranean basin in the 1990s \cite{FAO1990Unasylva162}. Among these, $57\%$ (17,056 events) exhibit a positive final AOI, defined as polygons with more than three vertices. These events, observed by FCI across multiple time steps and multiple pixels, are suitable for the calculation of FROS.

\begin{figure}[ht]
    \centering
    \includegraphics[width=.7\linewidth]{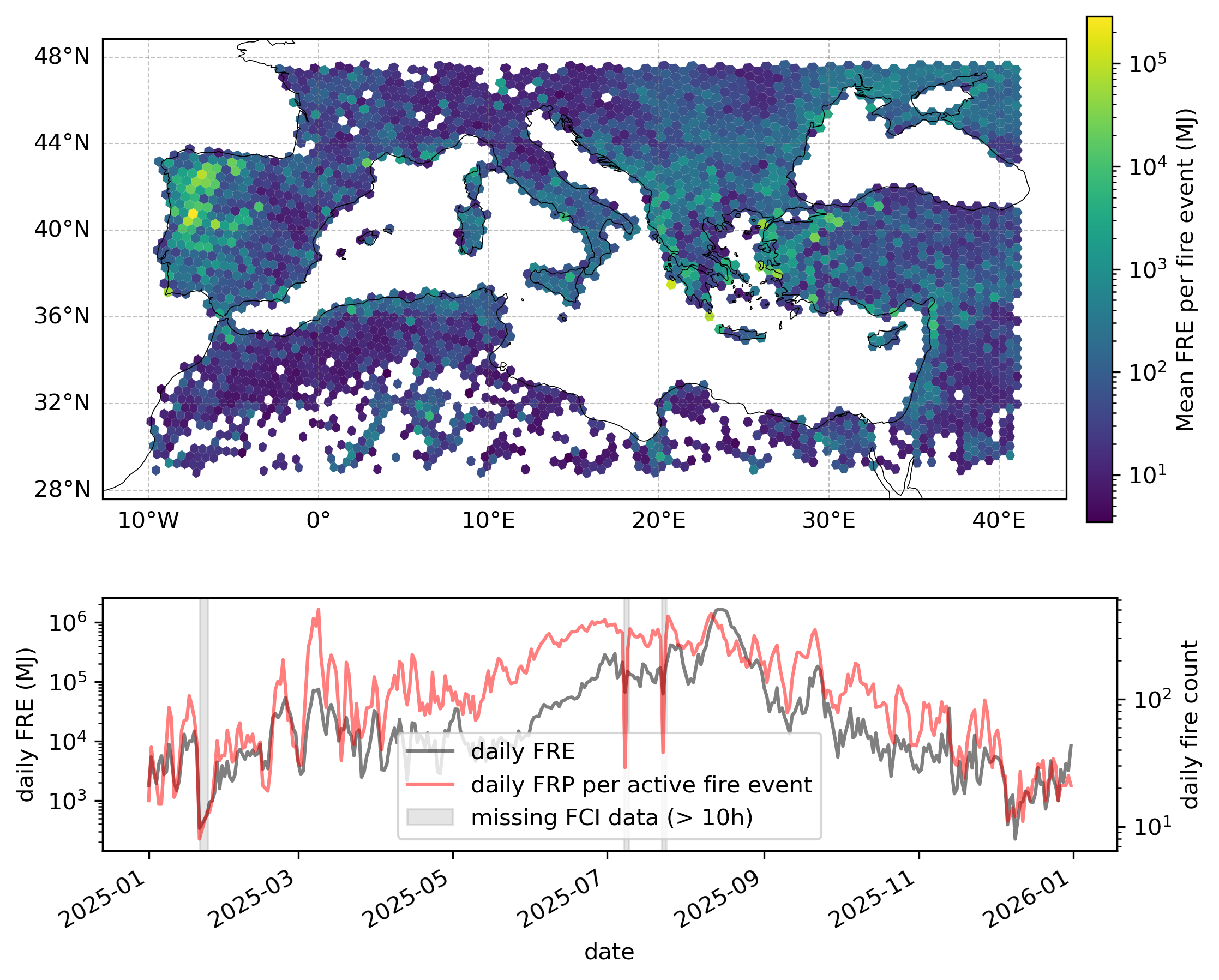}\
    \put(-345,288){(a)}
    \put(-345,122){(b)}
    \caption{Panel (a) shows the yearly sum of the Fire Radiative Energy (FRE) for all fire event of 2025 within the MED domain. The sum is computed for areas of $177~$ha (level 4 of h3 \cite{uber_h3_2018}. Panel (b) shows the daily  time series of FRP and total number of fire event over the MED domain for the year 2025.}
    \label{fig:MED3_fre}
\end{figure}

\begin{figure}[ht]
    \centering
    \includegraphics[width=.8\linewidth]{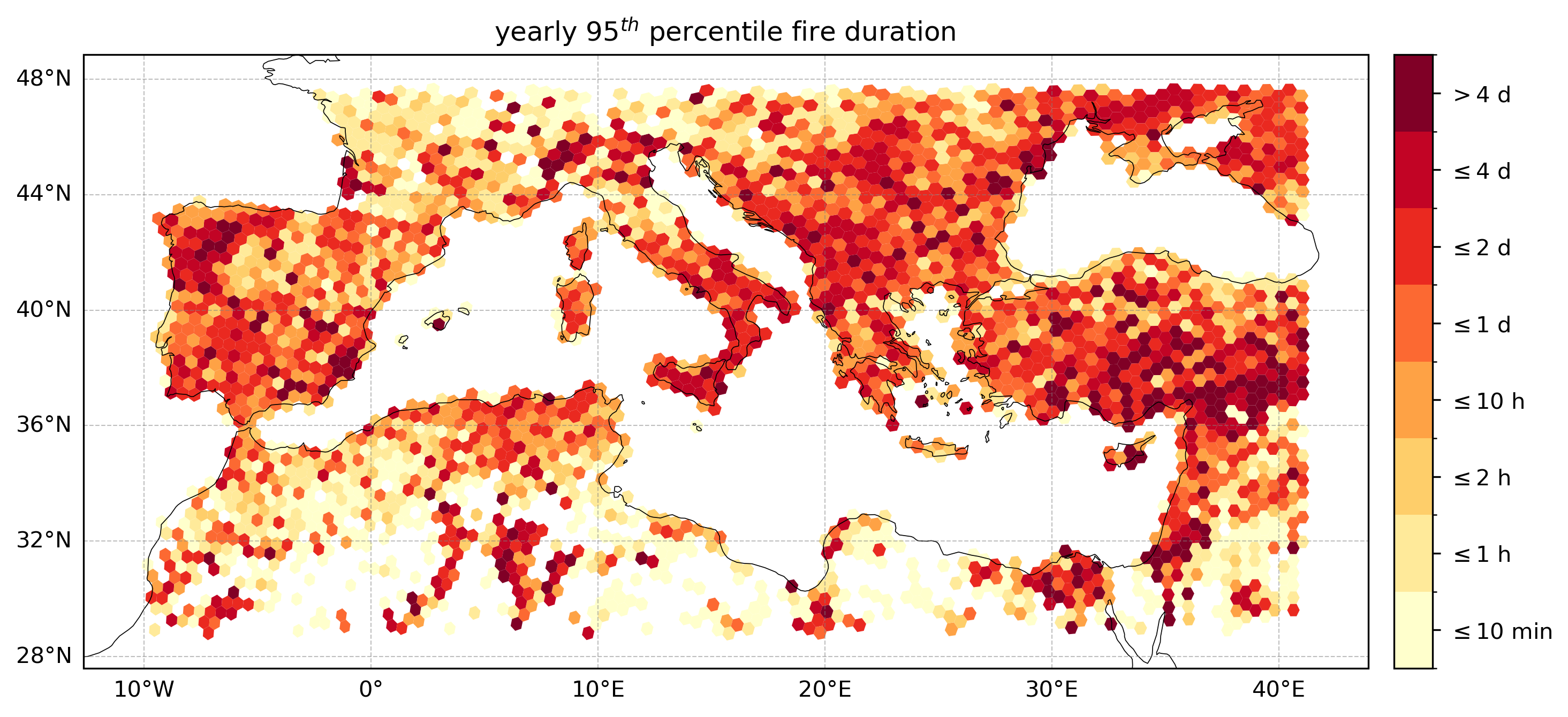}
    \caption{Map of the 95th percentile of the fire event duration of all fire events from 2025 within the MED domain. The 95th percentile is calculated for areas of $177~$ha (level 4 of the h3 grid \cite{uber_h3_2018}).}
    \label{fig:MED3_duration}
\end{figure}

\begin{figure}[ht]
    \centering
    \includegraphics[width=.8\linewidth]{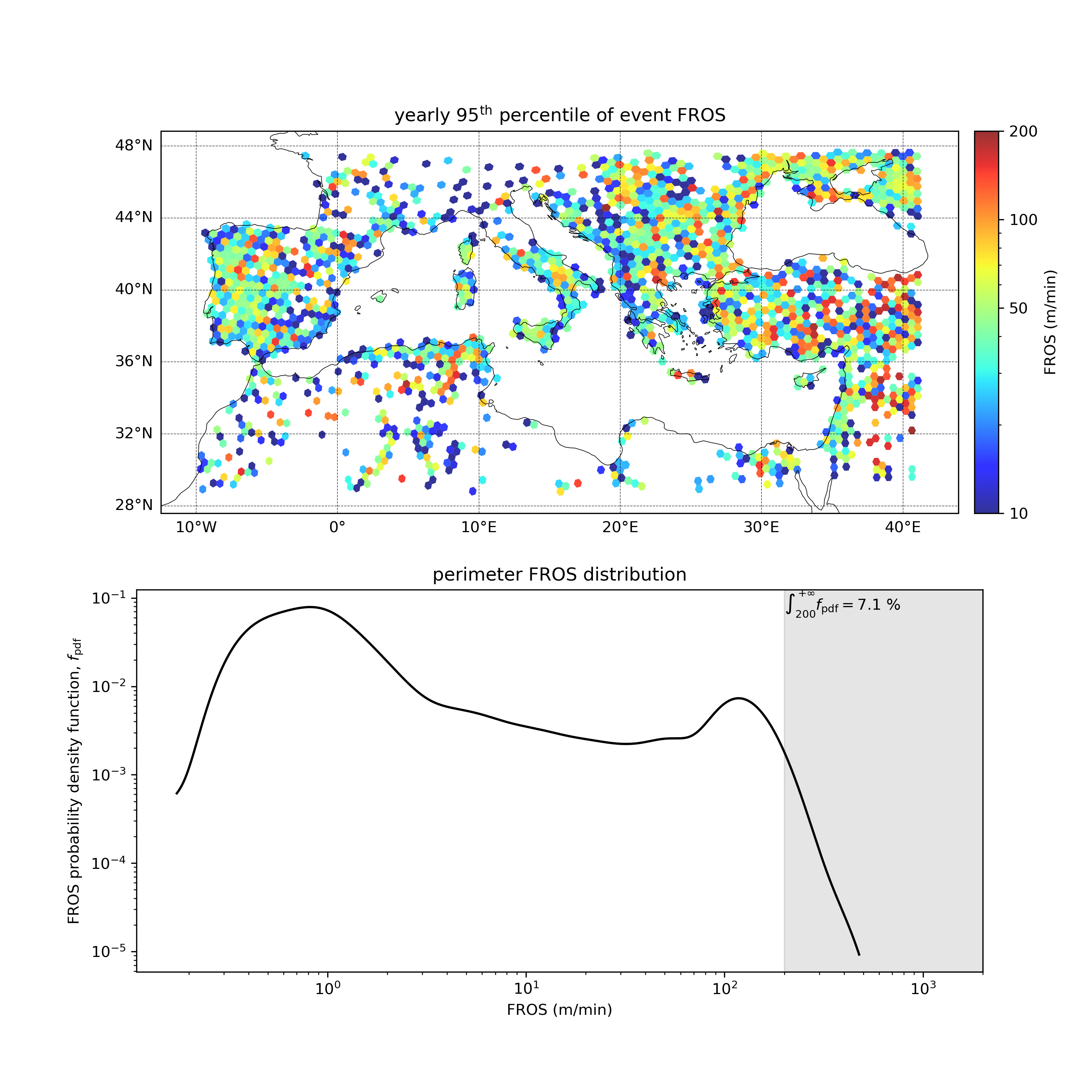}
    \put(-390,375){(a)}
    \put(-390,200){(b)}
    \caption{Panel (a) shows Map of the 95th percentile of the event FROS for all fire events from 2025 within the MED domain. The event FROS is calculated as the mean of the valid perimeter FROS of each perimeter of the event. The 95th percentile of the event FROS is calculated on areas of $177~$ha (level 4 of the h3 grid \cite{uber_h3_2018}). Panel (b) shows the distribution of all the perimeter FROS of all the events of 2025 within the MED domain. FROS perimeter above 200 m/min are not considered valid, and not used in the computation of the event FROS.}
    \label{fig:MED3_fros}
\end{figure}

This dataset provides an initial database describing the evolution of fire behavior events at 10-minute intervals, over a domain scale that is representative of the fire regime characteristic of the Mediterranean basin. The aim of this manuscript is not to carry out a fire regime analysis, but instead to illustrate the output of the FET algorithm and potential limitation of the current hotspot product. Figures \ref{fig:MED3_fre}, \ref{fig:MED3_duration} and \ref{fig:MED3_fros} present maps over the Mediterranean basin for 2025 of the yearly mean FRE per fire event, the 95th percentile of the fire event duration and the 95th percentile of the event FROS. The event FROS is calculated as the mean of all valid perimeter FROS for each event.
The daily FRE time series in Fig.~\ref{fig:MED3_fre}.b reveals a pronounced enhancement in fire activity during August 2025, predominantly driven by a major fire episode in the northwestern sector of the Iberian Peninsula, which is also evident in the mean FRE spatial distribution presented in Fig.~\ref{fig:MED3_fros}.a. This peak in fire activity is clearly discernible in the FRP time series in mid‑August and is accompanied by a nearly constant, slightly decreasing number of active fire detections. This pattern suggests an intensification of fire behavior, rather than a mere increase in the frequency of fire occurrences. The northwestern Iberian event was indeed characterized by the manifestation of several extreme fire episodes \cite{Sanchez2025}. By contrast, other intervals displaying an increase in daily FRE—such as late February, when fire activity was dominated by coordinated agricultural burning in southwestern France—are associated with substantial increases in the number of detected fire events.

In the same sector of northwestern Iberia, the map of mean fire duration (Fig. \ref{fig:MED3_duration}) reveals the occurrence of long-lasting fire events (exceeding 2 days). However, extended event duration is not exclusively associated with extreme episodes. Such persistent fires are also identified for example in the Central and Eastern Europe, central Turkey, and Sicily.

To complement this characterization of fire activity within the MED domain, Fig.~\ref{fig:MED3_fros} presents the 95th percentile of the event FROS on the same grid as in Figure \ref{fig:MED3_fre} and \ref{fig:MED3_duration}.
The distribution of the perimeter FROS in Fig \ref{fig:MED3_fros}.b shows two peaks of FROS values around $1$ and $100~$m/min which corresponds to the propagation of 1 pixel ($1~$km resolution) in 1 time period observation ($10~$min). Values above 200 m/min ($7\%$ of the perimeter FROS) are unrealistic \cite{Benali2023} and are removed from the calculation of the event FROS (non valid FROS).
The perimeter FROS calculation presented in the appendices fails to provide a result for approximately $30\%$ of the events with a final positive AOI. These failures are attributable to event with complex perimeter geometries that do not satisfy the constraints underlying the FROS algorithm.

\begin{figure}[!ht]
    \centering
    \includegraphics[width=.6\linewidth]{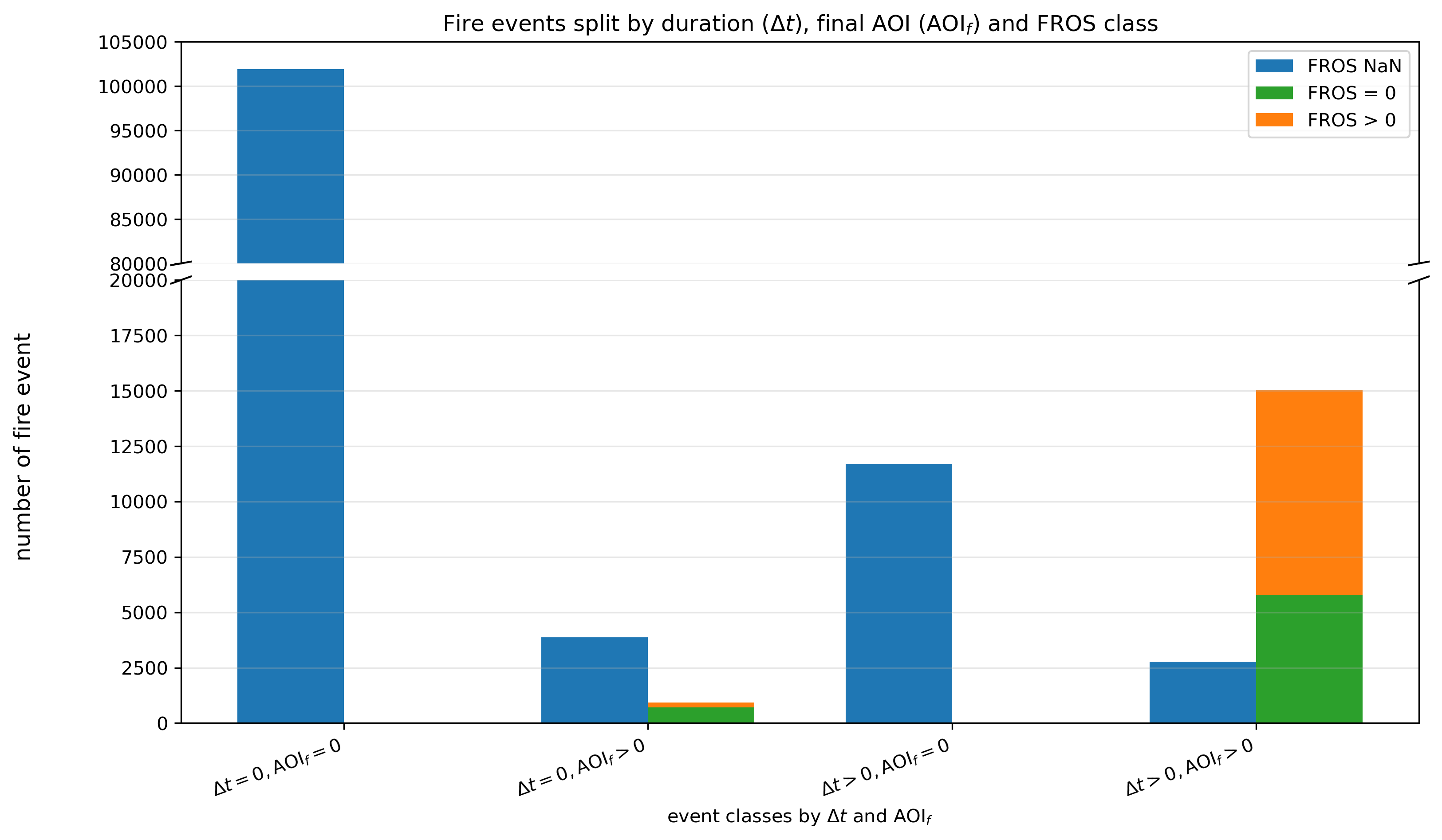}
    \caption{Forward Rate Of Spread (FROS) classes (no value, null, and positive) distribution across fire event categories defined by their spatiotemporal observation characteristics, specifically according to whether event duration and/or the final AOI is null or positive.}
    \label{fig:MED3_FROS_Class}
\end{figure}

Among the events for which a FROS is successfully computed, %($\text{FROS}\geq0$),
$57\%$ (8,761 events) are associated with a positive event FROS (i.e., propagating fires). Figure~\ref{fig:MED3_fros} presents the spatial distribution of these propagating fire events, while Figure~\ref{fig:MED3_FROS_Class} presents the distribution of the FROS classes (no value, null, and positive) across fire event categories. These categories are defined based on their spatiotemporal observation characteristics, specifically according to whether the event duration and/or the final AOI is null or positive. A substantial proportion of fire events exhibit zero duration and a null final AOI.
For these events, FROS computation predominantly fails. This limitation is intrinsic to the spatial resolution of the FCI sensor and is likely further exacerbated by false positives in the hotspot product that pass through the fixed hotspot mask filter present in the FET algorithm.

Within the MED domain, the spatial distribution of propagating fires indicates a higher occurrence of fast-propagating fires in the eastern part of the region. Turkey in particular exhibits a marked presence of fast-propagating fires, consistent with previous observations for 2024 \cite{Douglas2025}.

\subsection{Application during the 2025 fire season in Portugal - [PRT]}
\label{sec:port}
%~~~~~~
The FET algorithm was executed sequentially using $10~$min integration time period and a $10~$min dumping interval. Adding the $20~$min data latency, FRP time series per fire event were updated $30~$min behind real time. At this stage in the development of the algorithm, neither the FROS computation nor the alpha-shape procedure for determining the corresponding concave polygon had yet been implemented. Instead, the convex polygon formed by the cluster of hotspot was used.

The 2025 fire season in Portugal shows a total burned area nearly doubling the one of the previous year despite a comparable number of fire events according to EFFIS data \cite{EFFIS2025Port}. For context, 2024 was already a high fire activity season, ranking at the 79th percentile of the burned area record since 2006. The 2025 season is therefore exceptional in terms of burned area and is characterized by the occurrence of larger fires.

%To help ANEPC action, the FET algorithm was deployed operationally in late July to deliver fire activity monitoring, providing time series of FRP for every fire event detected by MTG-I1 over Portugal. An operational prototype website was developed for this purpose on a server made available by the Virtual Operation Support Team (VOST) association (see supplementary material).

\begin{figure}[ht]
    % First row: RGB
    \begin{subfigure}{0.22\textwidth}
        \includegraphics[width=\linewidth]{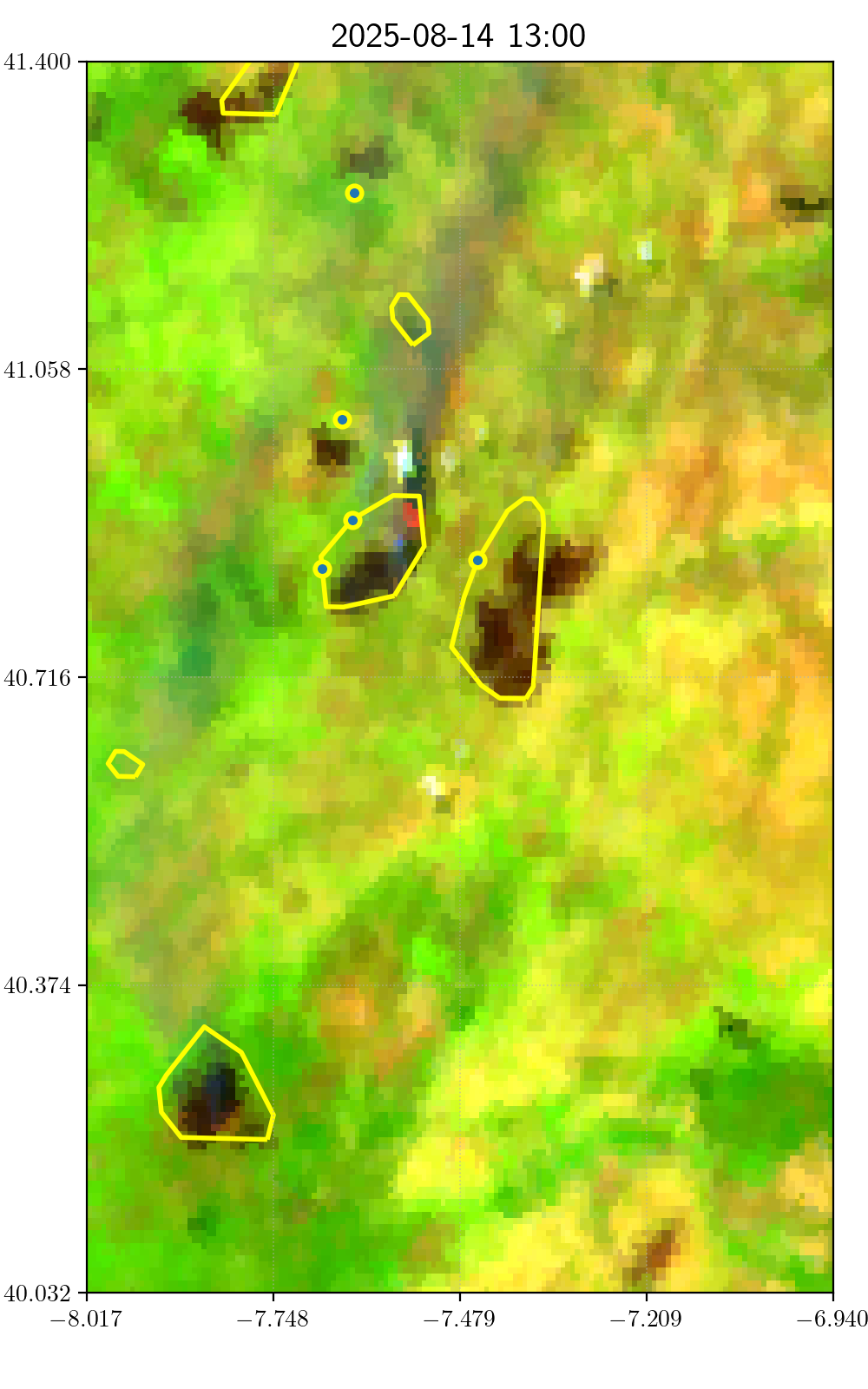}
    \end{subfigure}%
    \begin{subfigure}{0.22\textwidth}
        \includegraphics[width=\linewidth]{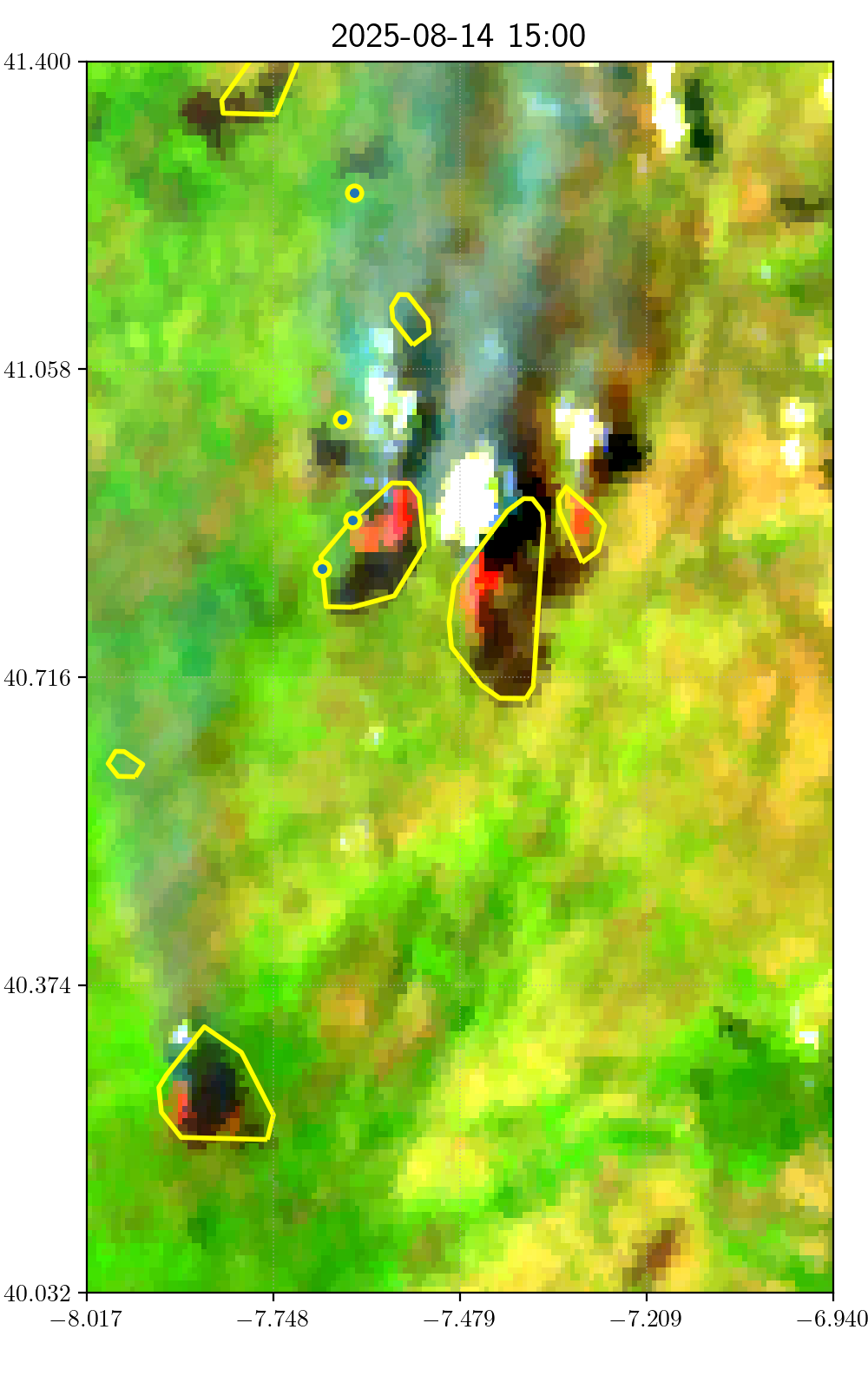}
    \end{subfigure}%
    \begin{subfigure}{0.22\textwidth}
        \includegraphics[width=\linewidth]{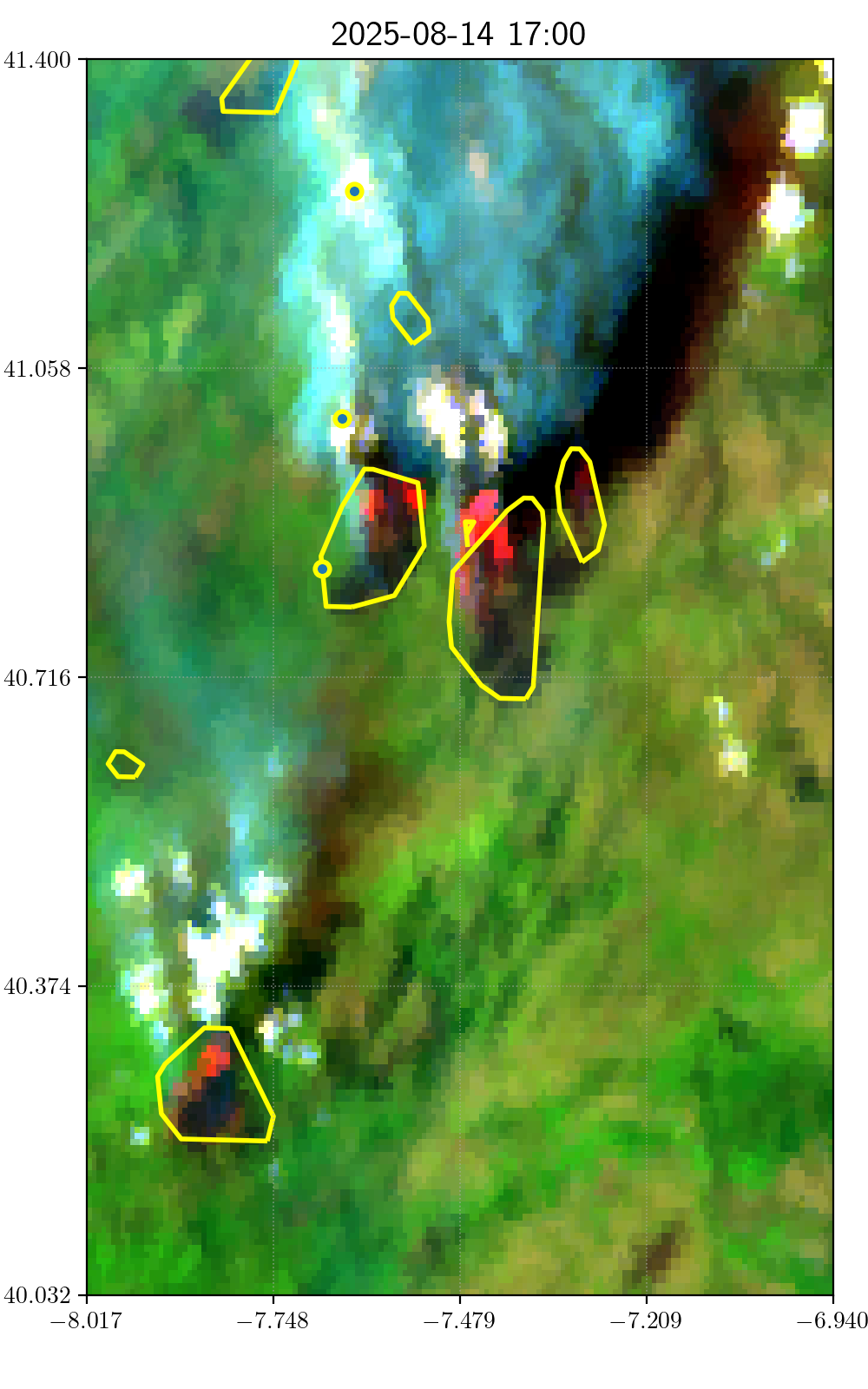}
    \end{subfigure}%
    \begin{subfigure}{0.22\textwidth}
        \includegraphics[width=\linewidth]{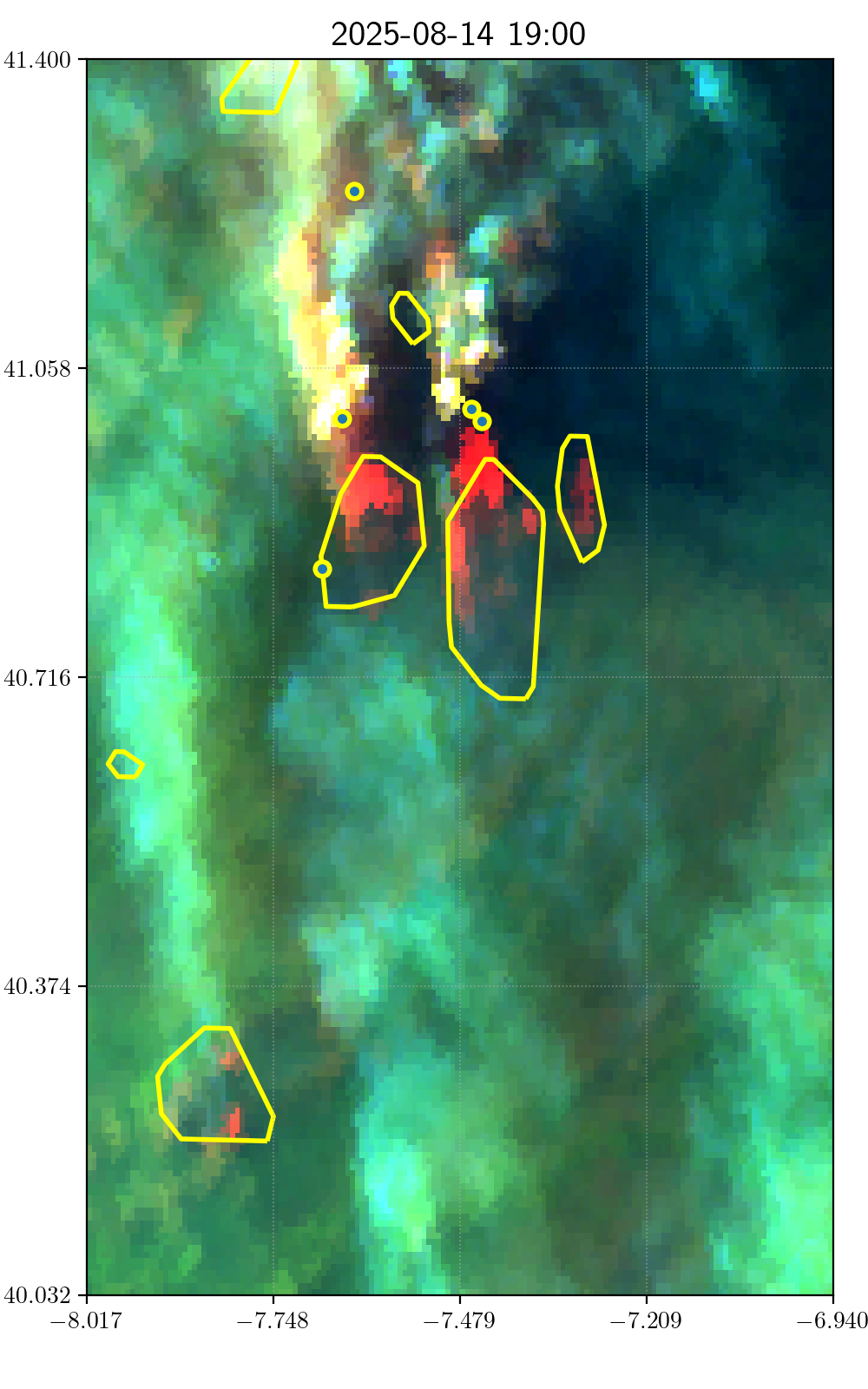}
    \end{subfigure}

    % Second row: IR
    \begin{subfigure}{0.22\textwidth}
        \includegraphics[width=\linewidth]{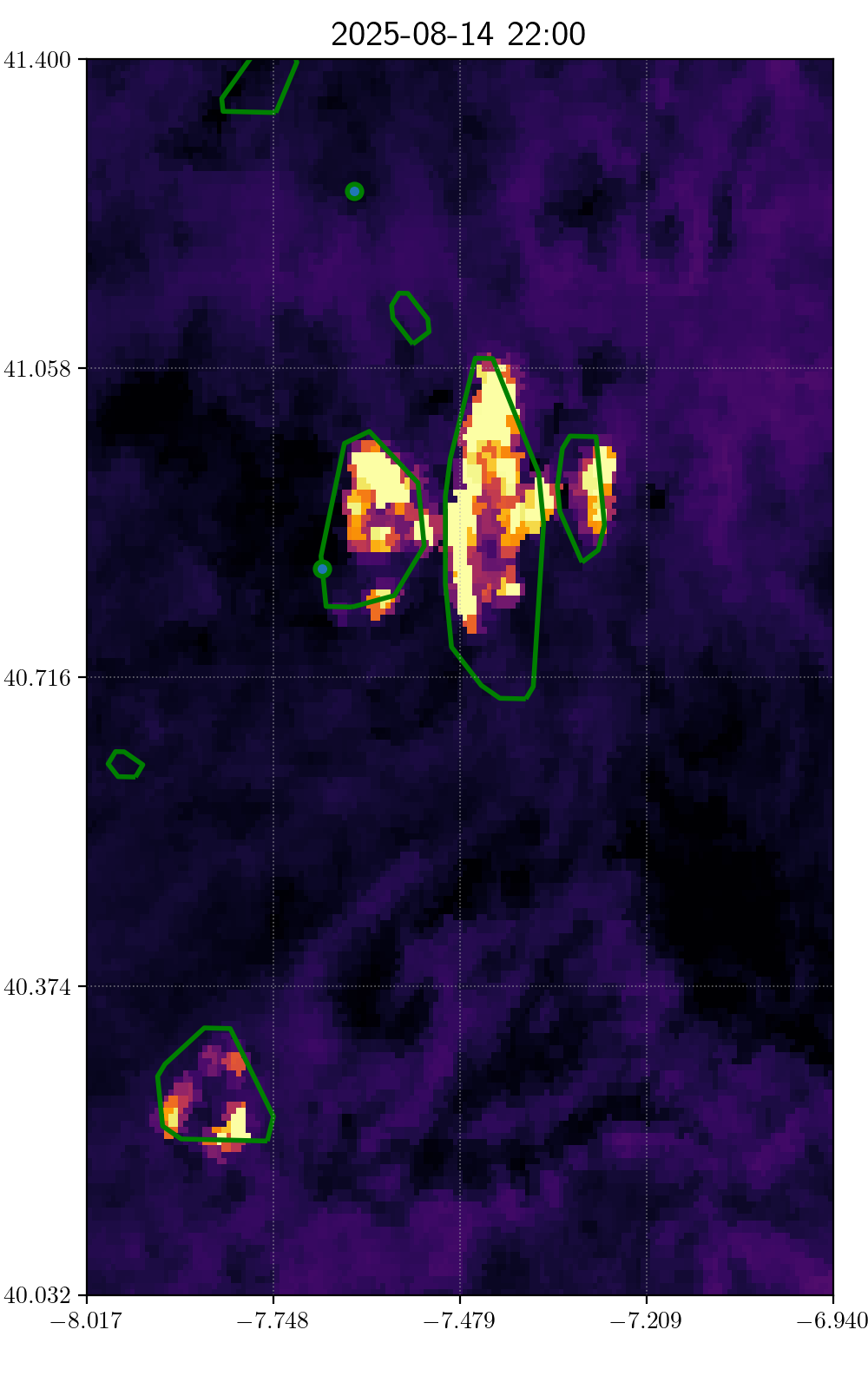}
    \end{subfigure}%
    \begin{subfigure}{0.22\textwidth}
        \includegraphics[width=\linewidth]{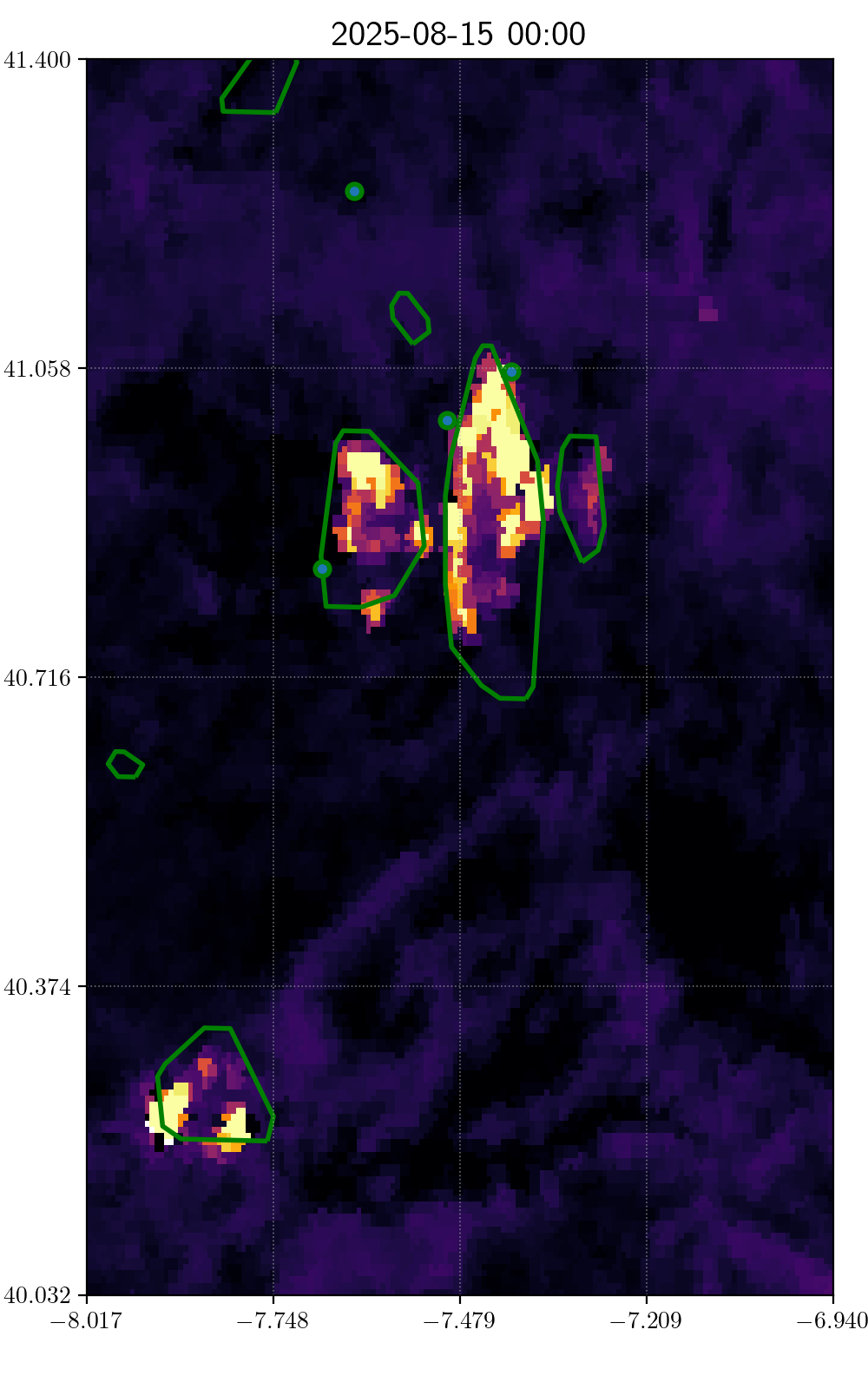}
    \end{subfigure}%
    \begin{subfigure}{0.22\textwidth}
        \includegraphics[width=\linewidth]{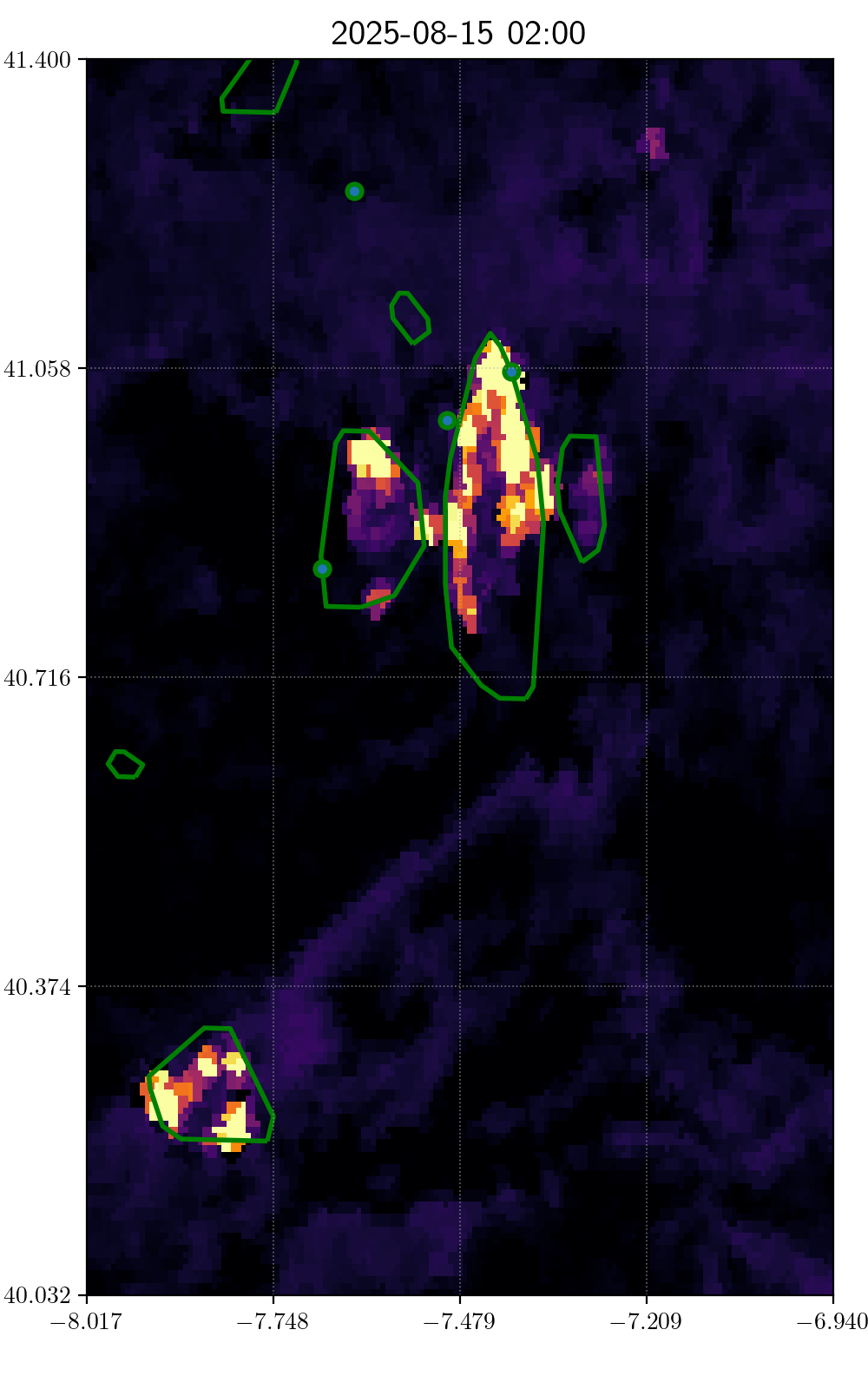}
    \end{subfigure}%
    \begin{subfigure}{0.244\textwidth}
        %\hspace{-2em} % shift image right
        \includegraphics[width=\linewidth]{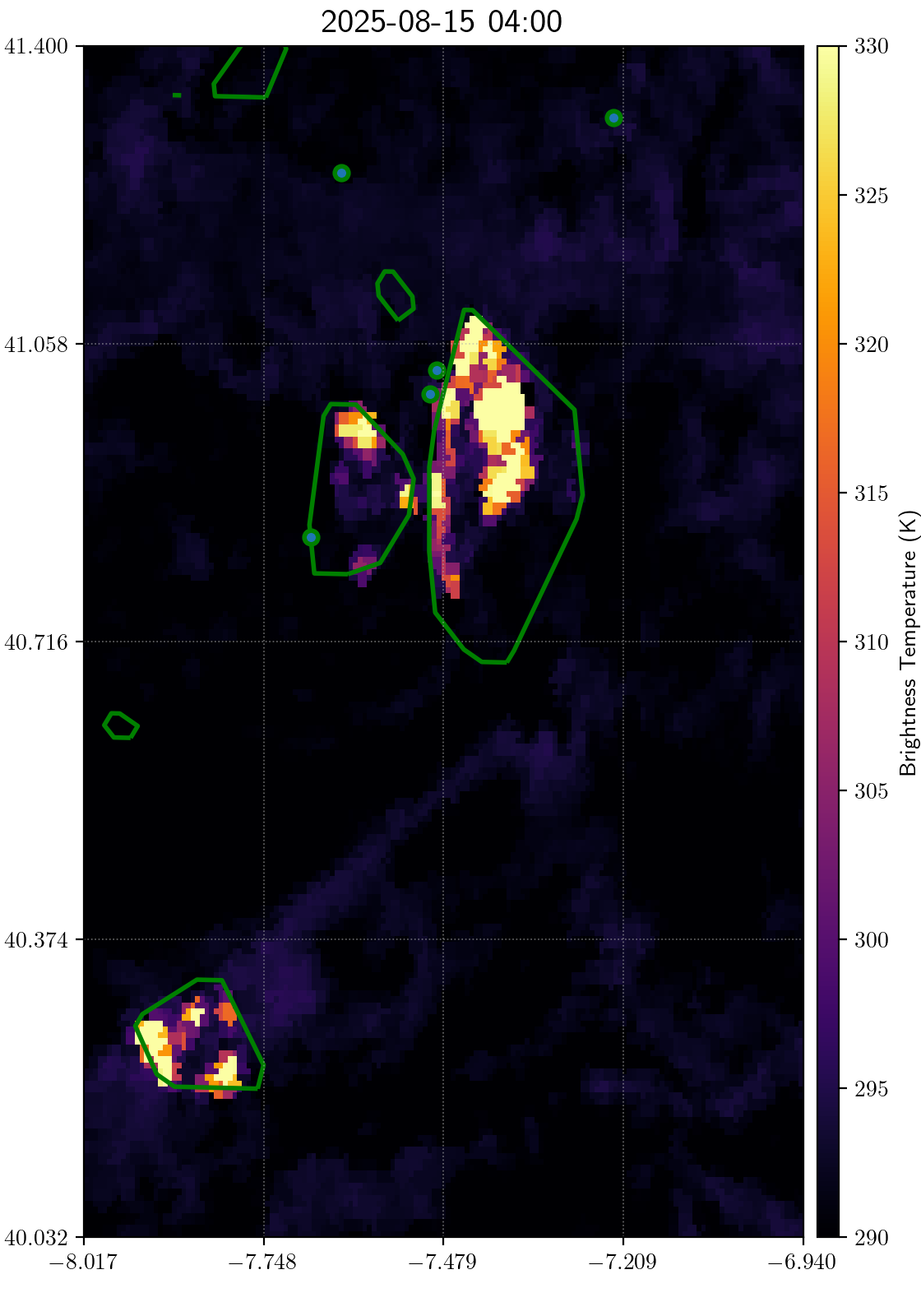}
    \end{subfigure}
    \caption{RGB (top) and Middle-Wave Infrared (MWIR, 3.8 µm; bottom) observations of fires in Portugal’s Região Centro, acquired by the FCI sensor onboard Meteosat Third Generation Imager-1 between 14 August 13:00 UTC and 15 August 04:00 UTC. RGB imagery corresponds to daytime hours, while MWIR data are shown for the period after sunset. Yellow and Green outlines (in top and bottom panels respectively) outlines indicate the fire areas of interest as delineated by the Fire-Event-Tracker algorithm.}
    \label{fig:FET_PROTUGAL_PRT}
\end{figure}

\begin{figure}[ht]
    \centering
    \includegraphics[width=\linewidth]{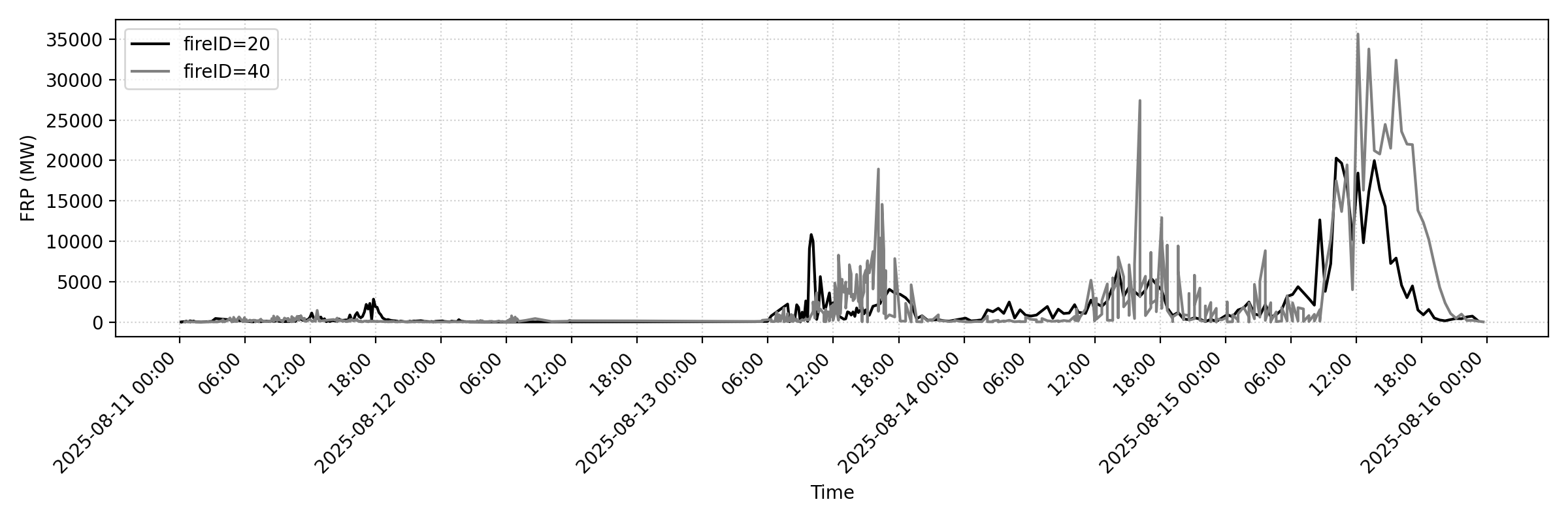}
    \caption{Time series of FRP for two fire events shown in Figure \ref{fig:FET_PROTUGAL_PRT}. The fire with ID $20$ corresponds to the event in the southwest corner of Figure \ref{fig:FET_PROTUGAL_PRT}, while the fire with ID $40$ is located at the center of the three merging fires in the northern part of the figure. These three fires merged on 15 August. }
    \label{fig:FET1D}
\end{figure}

Figure \ref{fig:FET_PROTUGAL_PRT} shows the evolution of FET computed AOI (grey polygon) for fires in center region of Portugal, between 14 August 13:00 UTC and 15 August 04:00 UTC, as well as FCI True color composite and Middle Wave Infra Red (MWIR, 3.8 $\mu$m) images. The website also reports time series of FRP for any fire event. Figure \ref{fig:FET1D} presents an example of FRP time series for two fire events depicted in Figure \ref{fig:FET_PROTUGAL_PRT}. The fire with ID $20$ corresponds to the event located in the southwestern corner of Figure \ref{fig:FET_PROTUGAL_PRT}, whereas the fire with ID $40$ is situated at the center of the cluster of three merging fires in the northern portion of the figure. With a latency of 30 minutes and an update frequency of 10 minutes, these data enable:

\begin{itemize}
    \item Identification of the location and spatiotemporal evolution of a pronounced re‑intensification of fire activity in the afternoon and night of 14 August. In particular, the difference in plume structure between 1300 and 1500 UTC on 14 August is clearly discernible. This transition in fire behavior is also evident in the time series shown in Figure \ref{fig:FET1D}, which displays an increase in Fire Radiative Power (FRP) for both fires $20$ and $40$. Fire $40$ exhibits a pronounced, abrupt increase in FRP at 1500 UTC.
    \item Reconstruction of fire progression during periods when other monitoring sources were unavailable. For example, the merging of the two eastern fires within the three-fire complex in the northern sector of Figure \ref{fig:FET_PROTUGAL_PRT} during the night of 14 August is captured. This merging is also reflected in the FRP time series of fire $40$, which shows an increase between 0000 and 0600 UTC on 15 August.
    \item Identification of actively burning areas for use in operational briefings, thereby supporting the reallocation of suppression resources. An example occurs at 1800 UTC on 16 August, during the end-of-day briefing.
\end{itemize}

\subsection{Application during the SILEX airborne campaign - [SILEX]}
\label{sec:SILEX}
\begin{figure}[!ht]
    \centering
    \includegraphics[width=\linewidth]{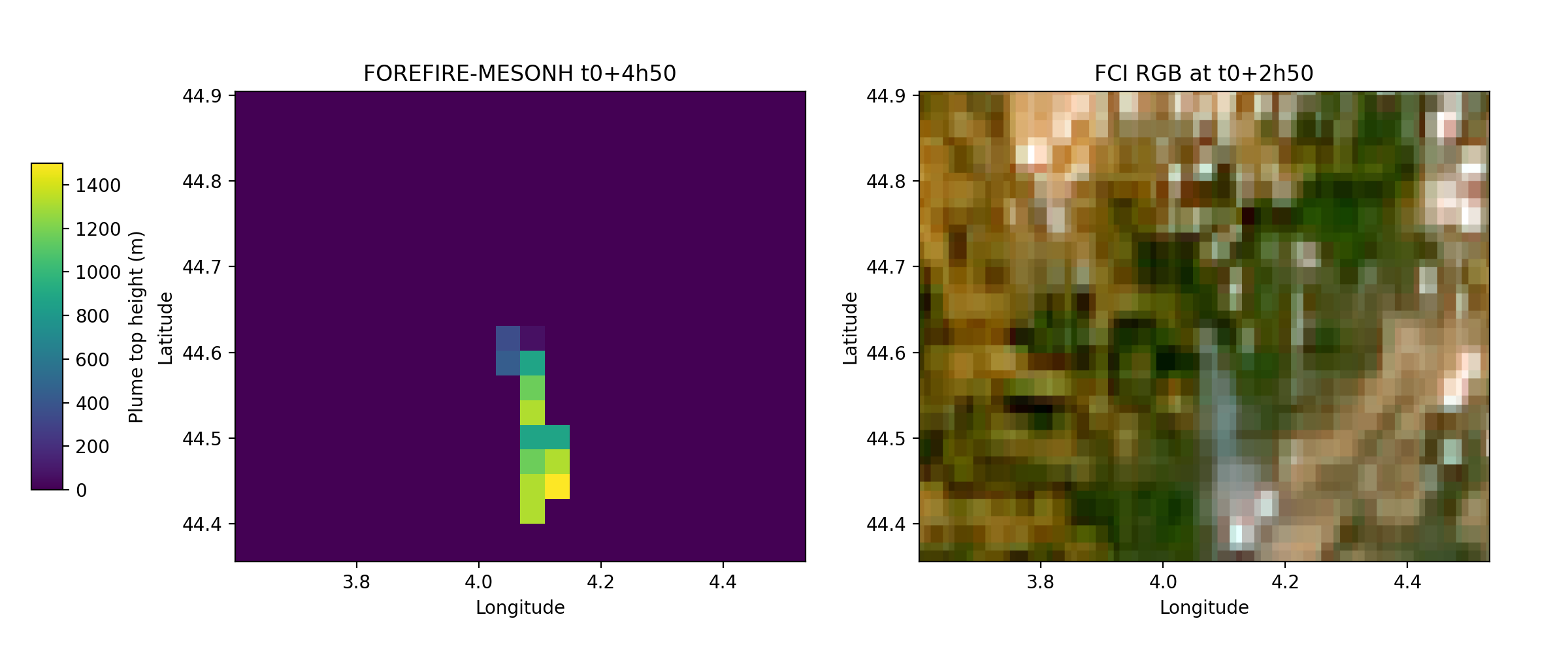} \\
    \includegraphics[width=\linewidth]{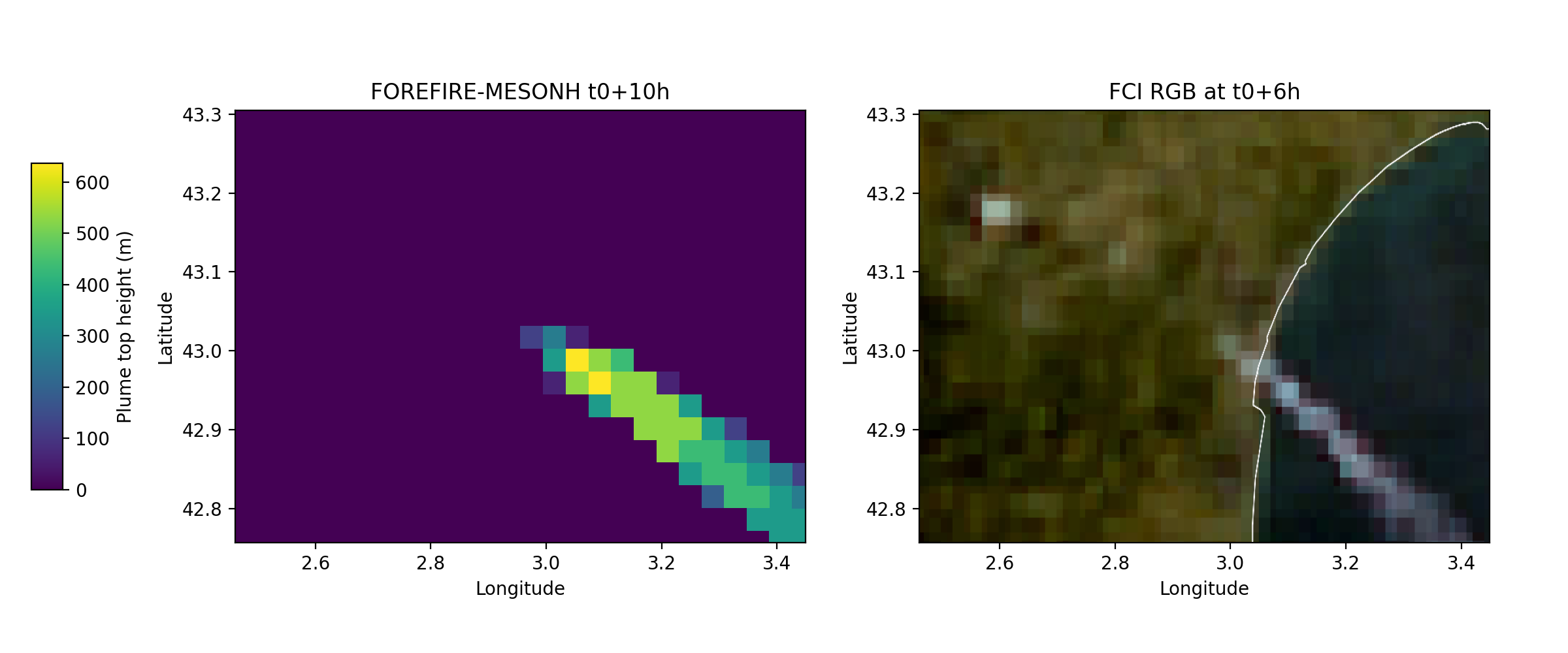}
    \caption{Comparison between FOREFIRE-MESONH plume simulation run within 1h after detection and FCI plume observation on two fires from the SILEX campaign where the ATR-42 was engaged: near Borne in Ardeche on 2025-07-17 (top row) and near Sijean in Aude on 2025-07-26 (bottom row). Plume top height from the FOREFIRE-MESONH simulation are reported. Time of the simulation are older than the times of observation: by 2h for Borne and 4h for Sijean.}
    \label{fig:ffmnh}
\end{figure}

%~~~~~~
During the SILEX campaign, the FET algorithm was deployed operationally on a server hosted by AERIS, that is expected now to host the official SILEX data repository in the near future. The SILEX campaign was the first deployment of the LSA-SAF hotspot product and the FCI data collection coupled with the FET algorithm. The output data were the same as for deployment in Portugal: fire AOI and FCI images (true color composite and MWIR) with a latency of 30 min and update every 30 min with a $10~$min time interval. The domain extended from west-Portugal to east-Italy and form south-Spain to north-France. This large domain was set to monitor fire activity in the south of France, but also any potential large scale plume transport form neighbored areas.
The AERIS server provided NRT monitoring of fire activity, allowing the identification of ignitions and early evolution to evaluate whether the ATR-42 should be engaged.
The decision process needed to be rapid, as the ATR-42 faced a four-hour delay for take-off authorization. Postponing the decision after the initial detection increased the likelihood of reaching a fire that was already under heavy suppression. To support this decision and provide the required input information on plume direction for inclusion in the flight plan authorization request, a fire-atmosphere coupled simulation was automatically initiated using FOREFIRE-MESONH \cite{Filippi2025} at the time of the first detection of a new fire event. Automatic simulations were run on the University of Corsica server, with a point ignition inferred from the first FCI detection.
There were two kind of possible automatic simulations, FIRSTGUESS (250.000 atmospheric grid points) and HD (2.200.000 atmospheric grid points), both with 2 nested domains with 50 levels. Both setup are run for 12 hours of forecast simulation; FIRSTGUESS has a domain extension and a resolutions of $800~$m and requires 50 CPUs for a 20 minutes wall time, while HD has a $160$ m resolution and requires 120 CPUs for a 1h wall time.
MesoNH domain is configured as a Large Eddy Simulation coupled to FOREFIRE at the highest resolution domain, with ECMWF data providing the lateral boundary conditions. Fuels were mapped at 10 m resolution on a 40km by 40km domain, while moisture content and live/dead partitioning followed regional defaults. Each simulation extended up to 12 hours lead time and produced fire spread isochrones, convective plume top height fields, smoke layer distributions, and near-surface concentration estimates. Plume top height is computed by finding the maximum altitude of a significant threshold of smoke tracer (0.1\% of the injection rate).
The automatic processing chain coupling FET and FOREFIRE-MESONH operates the FIRSTGUESS configuration, according to the following time line assuming FCI detect the fire at the first acquisition and that we required a transit time to reach the fire location of around 1h:
\begin{itemize}
  \item \textbf{T0}: Fire ignition.
  \item \textbf{T0+20 min}: First FCI active fire detection (hotspot) over the incident region and associated FRP are downloaded.
  \item \textbf{T0+30 min}: FET is run on the last 10 min time step. Fire event is created and ignition location is passed to FOREFIRE-MESONH. (The $+10~$min since download is depending on the number of fires to be processed. $10~$min corresponds to a worst case scenario.)
  \item \textbf{T0+30 min}: Forecast job is submitted. A first step of pre-processing starts where the domain and input data are configured (domain, fuels), ignition proxy is extracted from FCI hotspot, and ECMWF forcing field are downloaded.
  \item \textbf{T0+40 min} The FOREFIRE-MESONH job is submitted.
  \item \textbf{T0+50 min}: First hourly coupled forecasts outputs available (at least 6 hours ahead, fireline isochrones, plume height, near\mbox{-}surface smoke).
  \item \textbf{T0+1h} = $T_{submission}$: The flight plan authorization request is send to the flight control authority.
  \item \textbf{T0+4h}: ATR-42 take-off. Take-off time is schedule for the ATR-42 to enter the fire perimeter at T$_{submission}+4h$.
  \item \textbf{T0+5h}: ATR-42 at fire location to start observation.
\end{itemize}
\begin{figure}[!ht]
    \centering
    \includegraphics[width=0.5\linewidth]{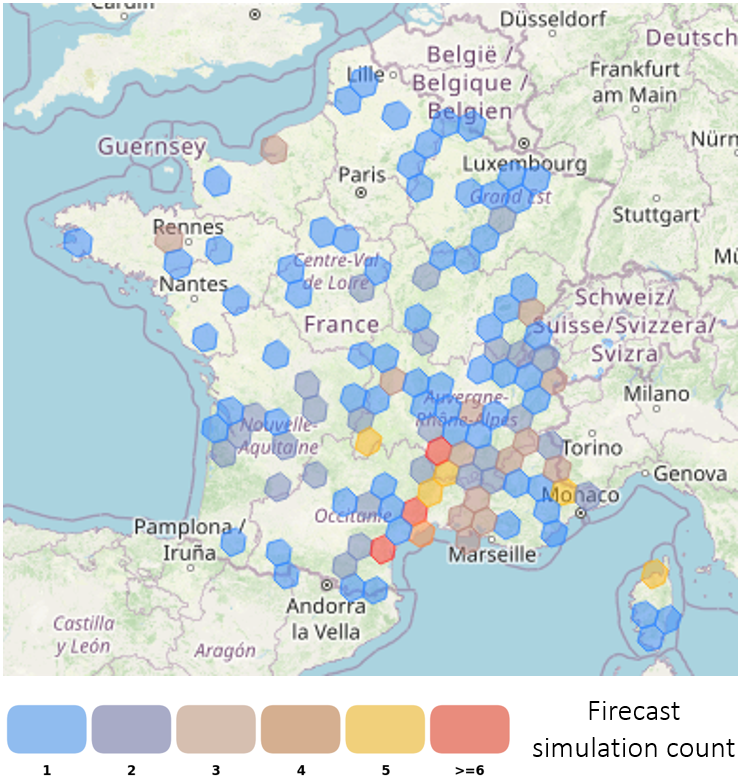}
    \caption{Spatial distribution of the 465 FCI triggered automated forecast launched for the SILEX Campaign, in blue more than 1 runs, in red, more than 20 runs.}
    \label{fig:fcastruns}
\end{figure}
Within one hour after detection we were able to estimate a plume direction using a fully automated processing chain. Over the three weeks of the SILEX campaign, 465 simulations were performed for each fire detection in France, with certain areas in Corsica and the south-west of the Rhône Valley showing a higher number of detections, as illustrated in Figure \ref{fig:ffmnh}.
Below, we present two cases in which the ATR-42 was deployed.
In both cases, the predicted plume reproduced the observed propagation direction as shown in Figure \ref{fig:ffmnh}, and helped to provide flights plan precise enough to support the flight campaign. However, the simulations showed a timing bias, with FOREFIRE-MESONH systematically lagging in capturing the onset of the initial updraft. This is most likely attributable to the relatively coarse grid spacing ($800$ m) used in the operational setup and the fact that the detection time may be well before the fire starting time. Since the main objective of the SILEX campaign was to provide plume direction forecasts to support flight planning, our analysis focuses on plume orientation within the first 10 h of simulation. For future developments, we plan to investigate the use of a nested high-resolution domain centered on the fire in order to better represent the early updraft dynamics.

\section{Discussion}
%-------------------
Conceived as a configuration-driven system, FET is capable of supporting both retrospective analyses and NRT operational applications. With regard to the latter, it has demonstrated the capability to process high-frequency input streams—such as the current 10-minute hotspot detections from FCI—even when applied over the domain defined by the SILEX configuration. FET supports continuous fire monitoring at a temporal resolution that far surpasses that of polar-orbiting sensors, while approaching the kilometer spatial resolution offered by several polar-orbiting instruments.

FET provides insight on fire behavior evolution from a vector perspective (this is not a gridded product) at $10~$min temporal resolution and at the scale of fire event. In the current version fire behavior metrics of FROS and FRP are computed for every events and are attached to the timely AOI that are defined from the spatio-temporal clustering of the LSA-SAF hotspot product.

\begin{figure}[!ht]
    \centering

    \includegraphics[width=0.9\linewidth]{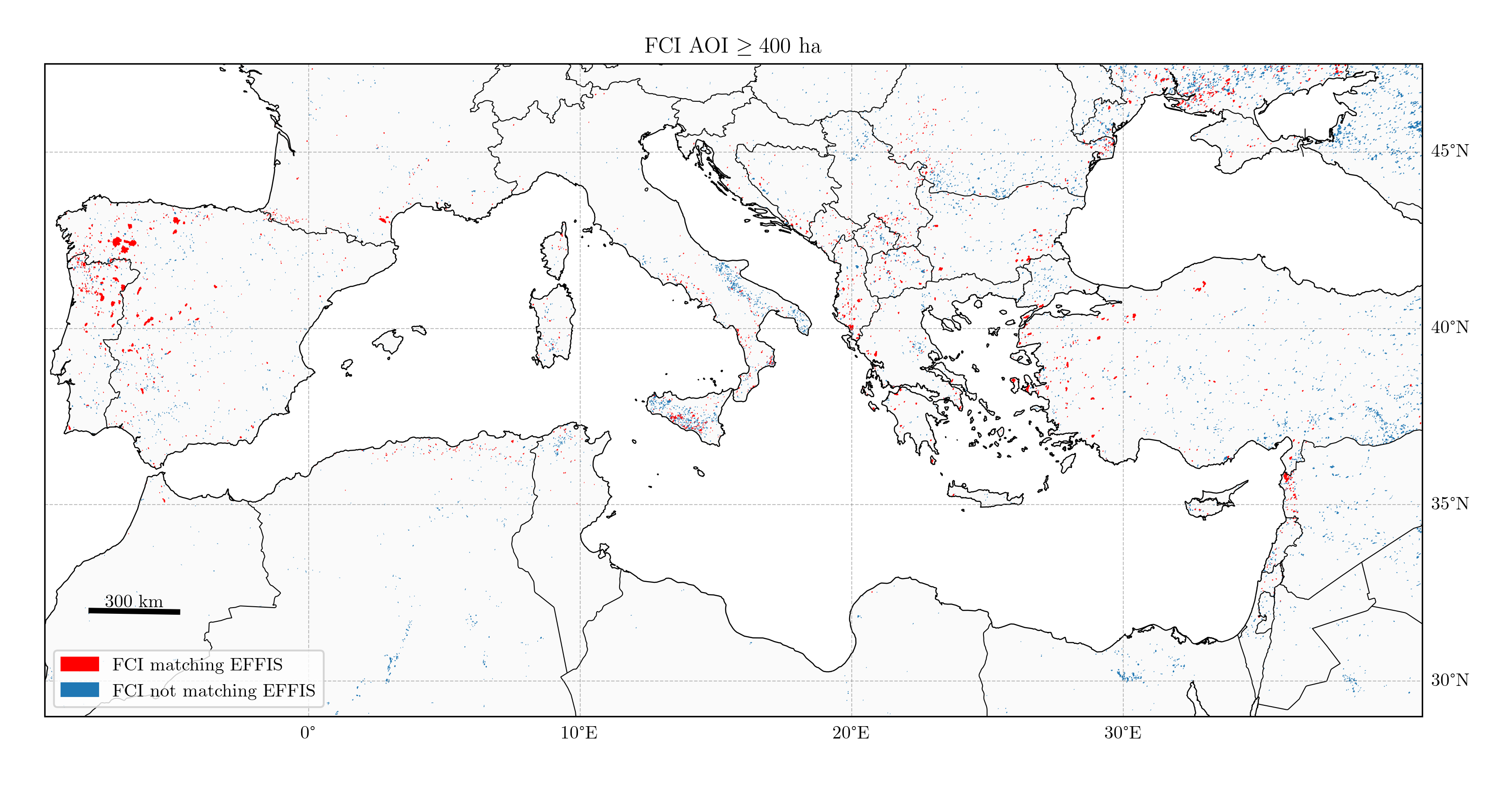} 
    \put(-460,235){(a)}
    \\
     \includegraphics[width=0.9\linewidth]{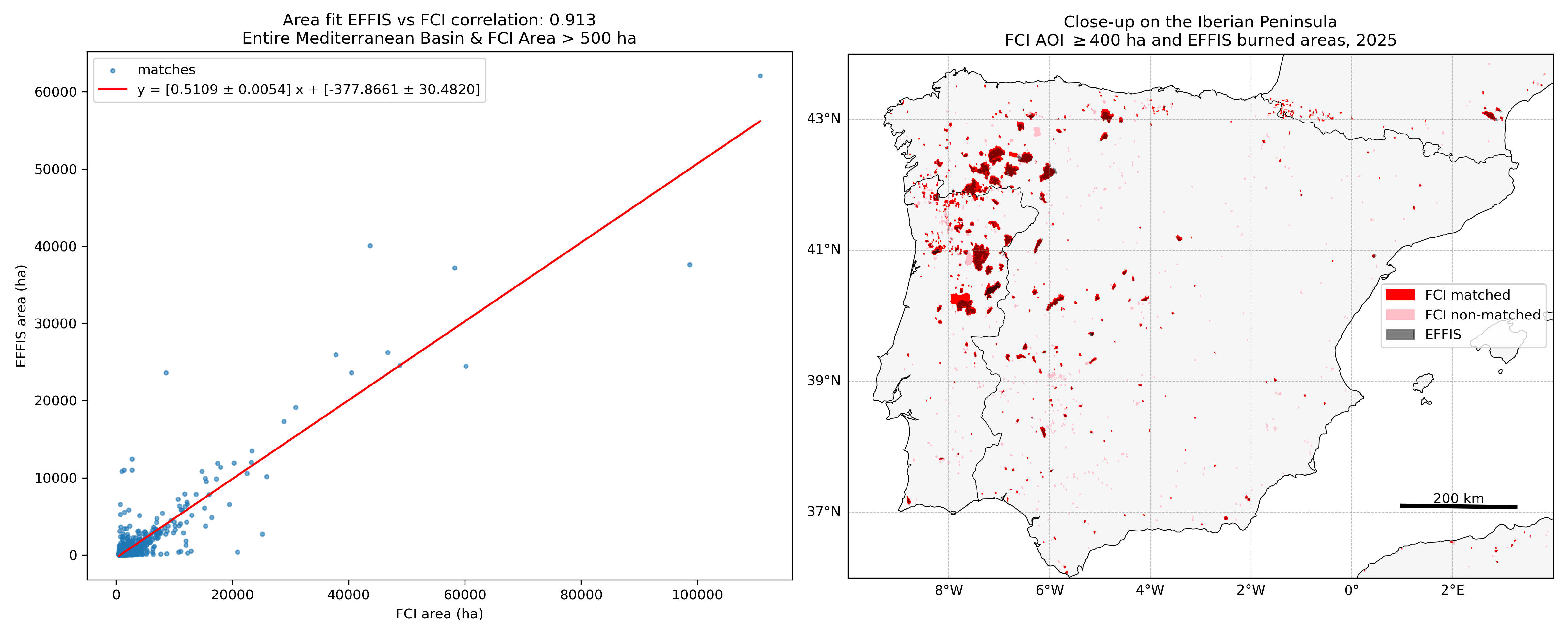}
    \put(-445,182){(b)}
    \put(-210,182){(c)}
    \caption{Comparison between the final Area of Interest (AOI) delineated by FET and the EFFIS Burned Area (BA) for the same 2025 fire events with AOI$>400~$ha. Panel (a) displays all 2025 fire events identified by FET from FCI hotspot detections that either match (in red) or do not match (in blue) an EFFIS BA. Panel (b) presents the correlation between EFFIS-BA and FET-AOI sizes. Panel (c) provides a magnified view of panel (a), focusing on the northern Iberian Peninsula where EFFIS BA polygons are overlaid in grey.}
    \label{fig:FET_EFFIS_MED}
\end{figure}

To assess how well AOI corresponds to the traditional BA product, Fig. \ref{fig:FET_EFFIS_MED} presents a comparison of the final FET AOI with the EFFIS BA dataset over the same temporal interval and spatial domain.
Small fires, whose geometry are less likely to be capture by FCI coarse spatial resolution are removed from the analysis. Only fire larger than $400~$ha are considered in Figure \ref{fig:FET_EFFIS_MED}; this counts 7,535 fire events of the 17,056 propagative fire event ($44\%$) introduce previously.

The threshold of $400~\text{ha}$ is justified in the appendices using a FROS performance analysis.

Figure \ref{fig:FET_EFFIS_MED}.a shows that the final AOI from FET is well correlated ($r=0.88$) with the EFFIS BA\cite{EFFIS2007}, with a roughly 2-fold difference. As expected, this correlation decreases for small fires, as EFFIS BA is based on satellite sensors with much higher resolution (250 m and lower \cite{EFFIS2007}) than FCI.

Figure \ref{fig:FET_EFFIS_MED}.b shows the spatial match between FET and EFFIS over the domain of the MED configuration, while Figure \ref{fig:FET_EFFIS_MED}.c is a close up of previous figure on the northern Iberian peninsula. the FET algorithm captures well all the large fires of the highly active 2025 fire season of the northern Iberian peninsula. Furthermore Fig \ref{fig:FET_EFFIS_MED}.c shows that there is no mismatch pattern between the two estimated areas.
When considering only AOI larger than $400~$ha, FET already counts more fires than EFFIS. Across the entire MED domain, $76\%$ of FET events are not reported in the EFFIS dataset. 
This percentage varies significantly depending on the region of interest. For instance, it decreases to 45\% within the spatial extent shown in Figure \ref{fig:FET_EFFIS_MED}.c.
The FCI hotspot product is returning a high rate of false positive detections. However, because the EFFIS BA product is not generated through a fully automated processing chain, it is challenging to fully characterise the origin and nature of the discrepancies observed in Figure \ref{fig:FET_EFFIS_MED}. 

The primary objective of this study is to demonstrate the capabilities of FET using the version of the FCI hotspot product from LSA-SAF that was available at the time of the analysis (January 2026). This product is less than one year old and remains under active development and refinement. 

Without explicitly attempting to filter out potential false positives at this stage, several aspects of the current results already provide valuable insights and directions for the future evolution and optimisation of FET.
\begin{itemize}
\item The total number of detected fire events (139,604) is very high and is likely to include a substantial proportion of occurrences associated with industrial facilities and/or solar farms. FET identifies numerous fire events in areas where the presence of natural vegetation is minimal, for instance in southern Algeria or eastern Syria. The fixed hotspot mask based on VIIRS is not effective enough to mask false positive in all regions of the Mediterranean basin. The method that was mostly tested on western European VIIRS hotspots for the SILEX airborne campaign needs to be improved to be applied to the whole MED domain. The use of the FCI 2025 archive will be investigated in next update.
\item Within a single country, regions characterized by a high proportion of well-matched AOIs may be directly adjacent to regions where the AOIs show poor correspondence (for example, see Sicily). This spatial heterogeneity may be associated with regional differences in fire regimes, such as presence of small-scale agricultural burns that are not recorded in the EFFIS database. A more detailed analysis of fire-event frequency and associated fire-behavior metrics is planned for 2025 in order to derive a more robust characterization of fire types.
\item Despite being based on a much lower sensor resolution than in the EFFIS approach, FET was still able to detect several fire events during the agricultural controlled burning episode that took place in the Pyrenees-Atlantiques region ($1^{\text{o}}\text{W}$,  $43^{\text{o}}\text{N}$) in late February 2025. .
\end{itemize}

In FET, the FRP value is calculated as the simple sum of the FRP assigned to each AOI within an event. To verify the timely geometric clustering underlying this summed FRP value, time series of FRP have been compared with an alternative method that relies on simpler bulk aggregation of all hotspots located within the event perimeter (personal communication Victor Penot).
FROS, on the other hand, is still in a development phase. Relying solely on the distance between the last two AOIs polygons (see methodology in the appendices) is very appealing in terms of computational efficiency. However, due to the low resolution of the FCI, this approach introduces substantial discontinuities or quantization in the FROS calculation (see $100~$m/min in Fig \ref{fig:MED3_fros}.b). The reason is that the displacement of the propagating front between two consecutive frames may be smaller than the pixel size in many situation. This first application of the FROS algorithm in the configuration of the MED domain shows 
that more time information is needed than that provided by the last two AOI polygons in order to overcome the lack of spatial information and avoid discontinuous FROS estimates.

\begin{figure}[ht]
    % First row: RGB
    \begin{subfigure}{0.22\textwidth}
        \includegraphics[width=\linewidth]{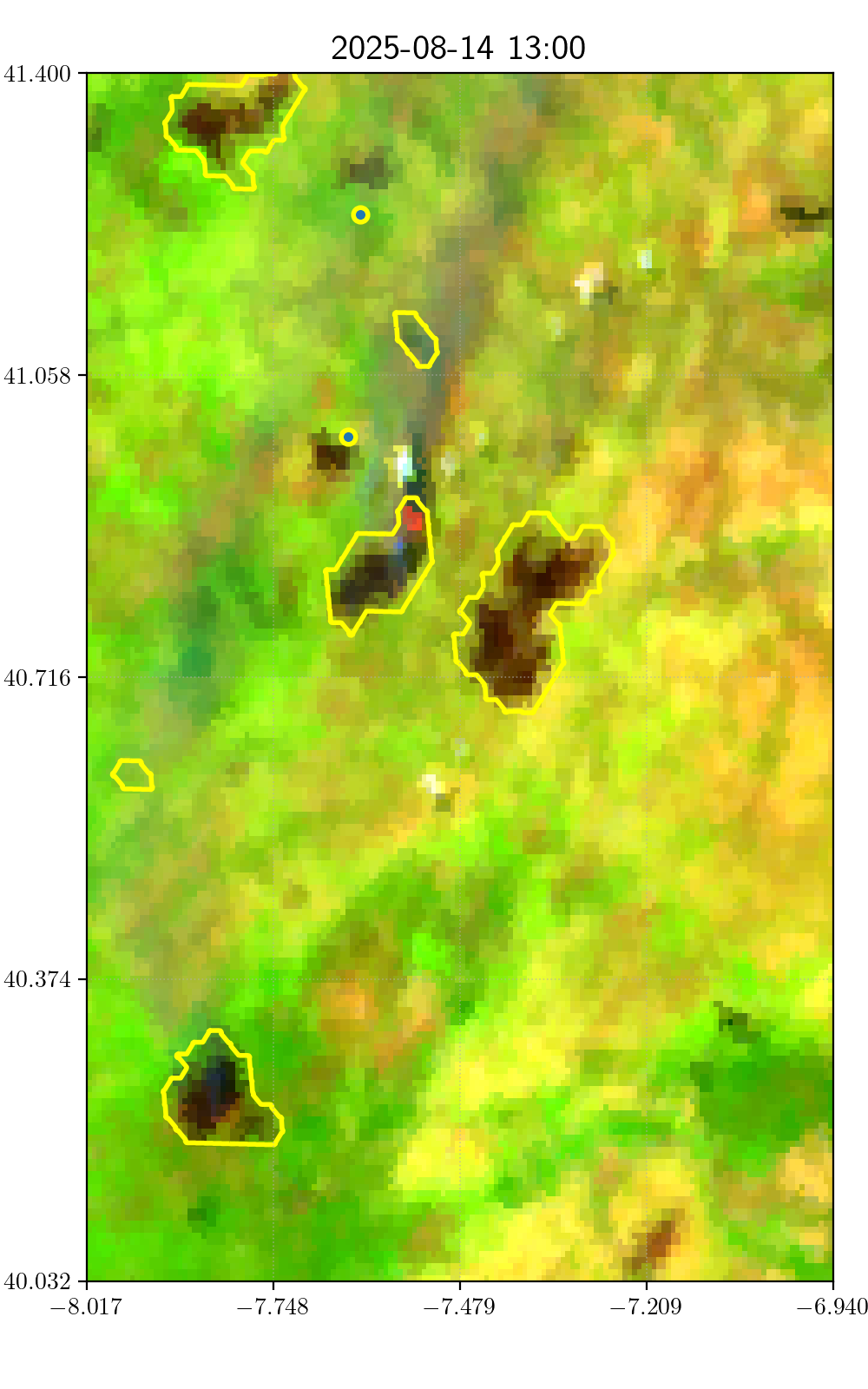}
    \end{subfigure}%
    \begin{subfigure}{0.22\textwidth}
        \includegraphics[width=\linewidth]{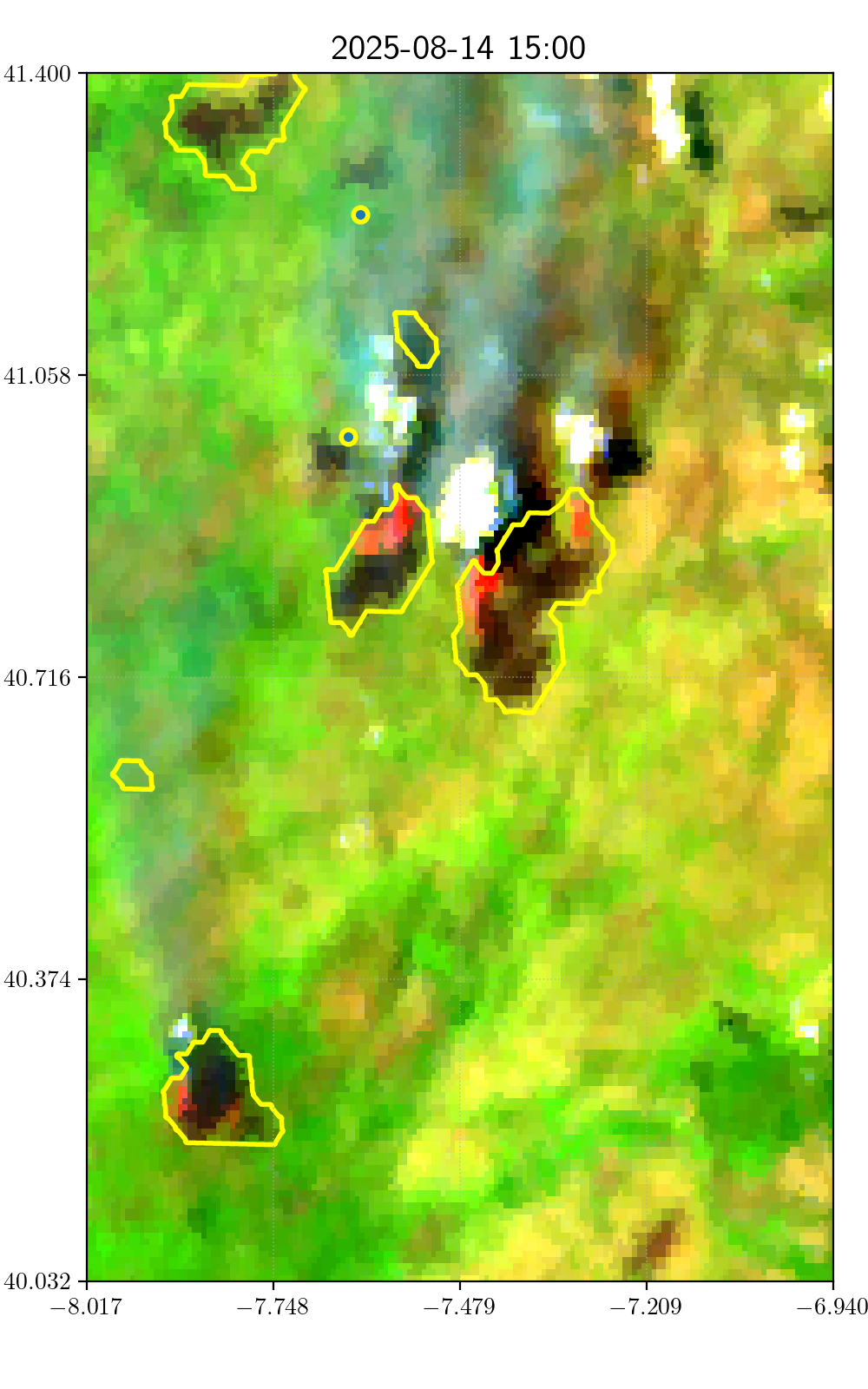}
    \end{subfigure}%
    \begin{subfigure}{0.22\textwidth}
        \includegraphics[width=\linewidth]{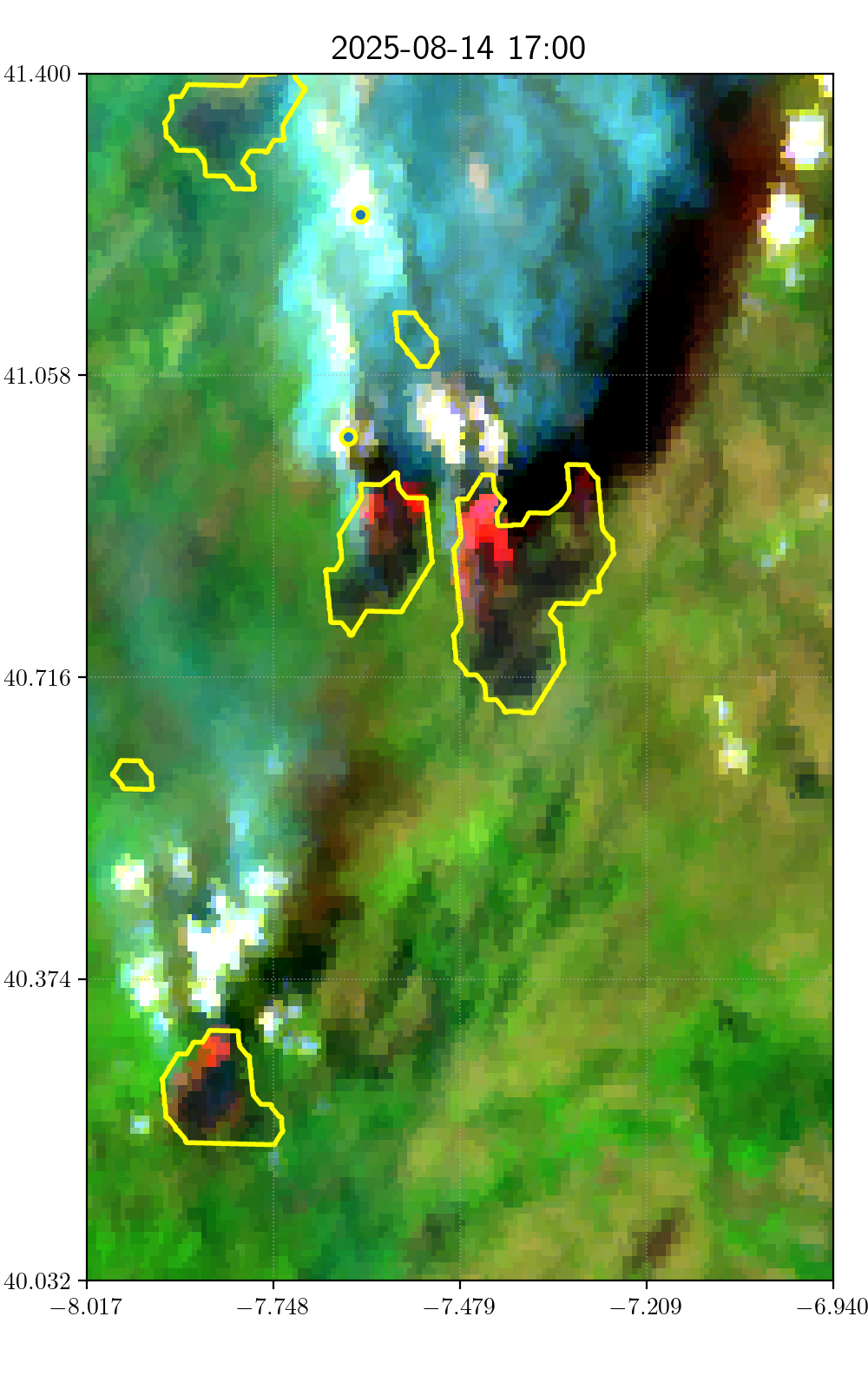}
    \end{subfigure}%
    \begin{subfigure}{0.22\textwidth}
        \includegraphics[width=\linewidth]{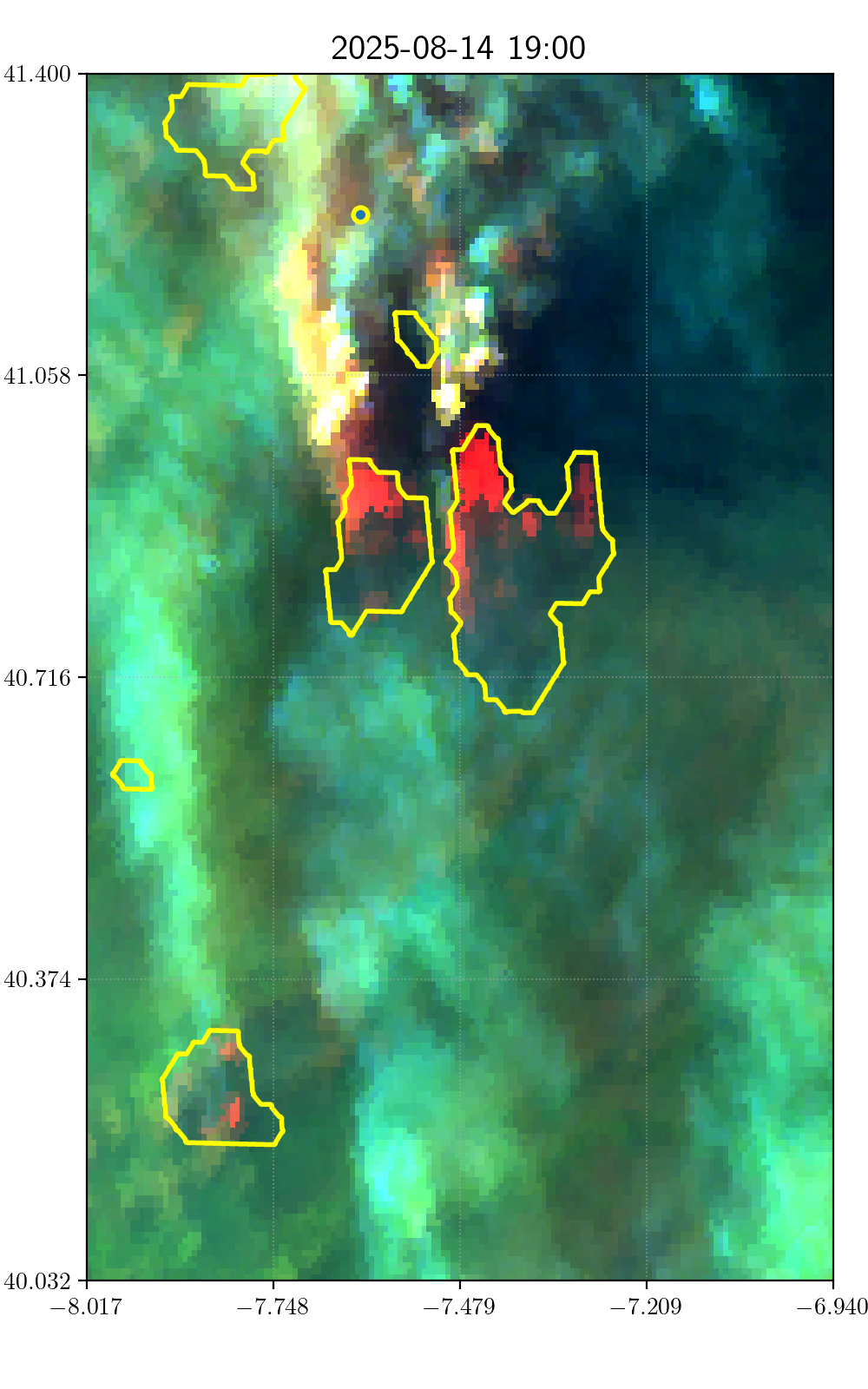}
    \end{subfigure}

    % Second row: IR
    \begin{subfigure}{0.22\textwidth}
        \includegraphics[width=\linewidth]{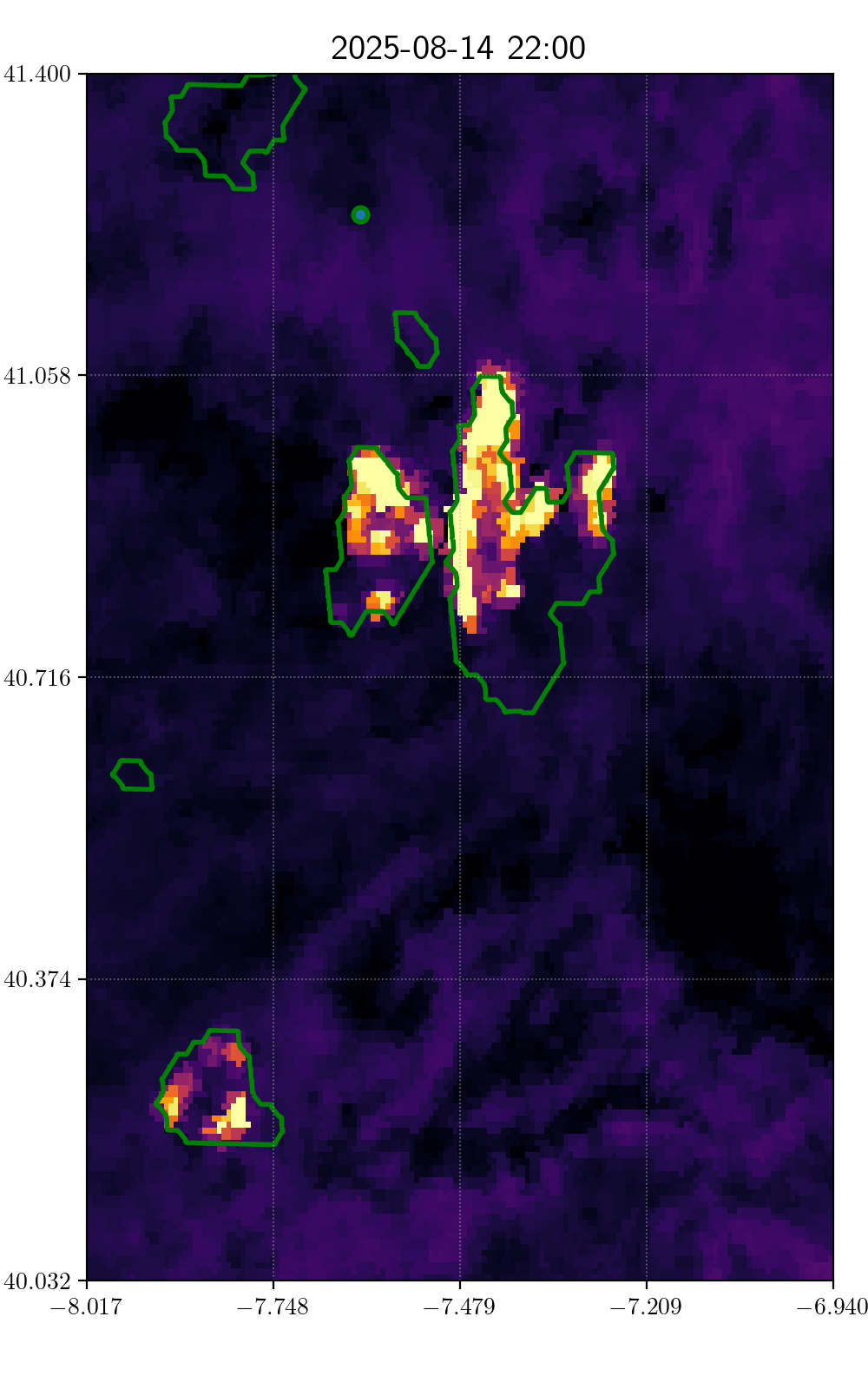}
    \end{subfigure}%
    \begin{subfigure}{0.22\textwidth}
        \includegraphics[width=\linewidth]{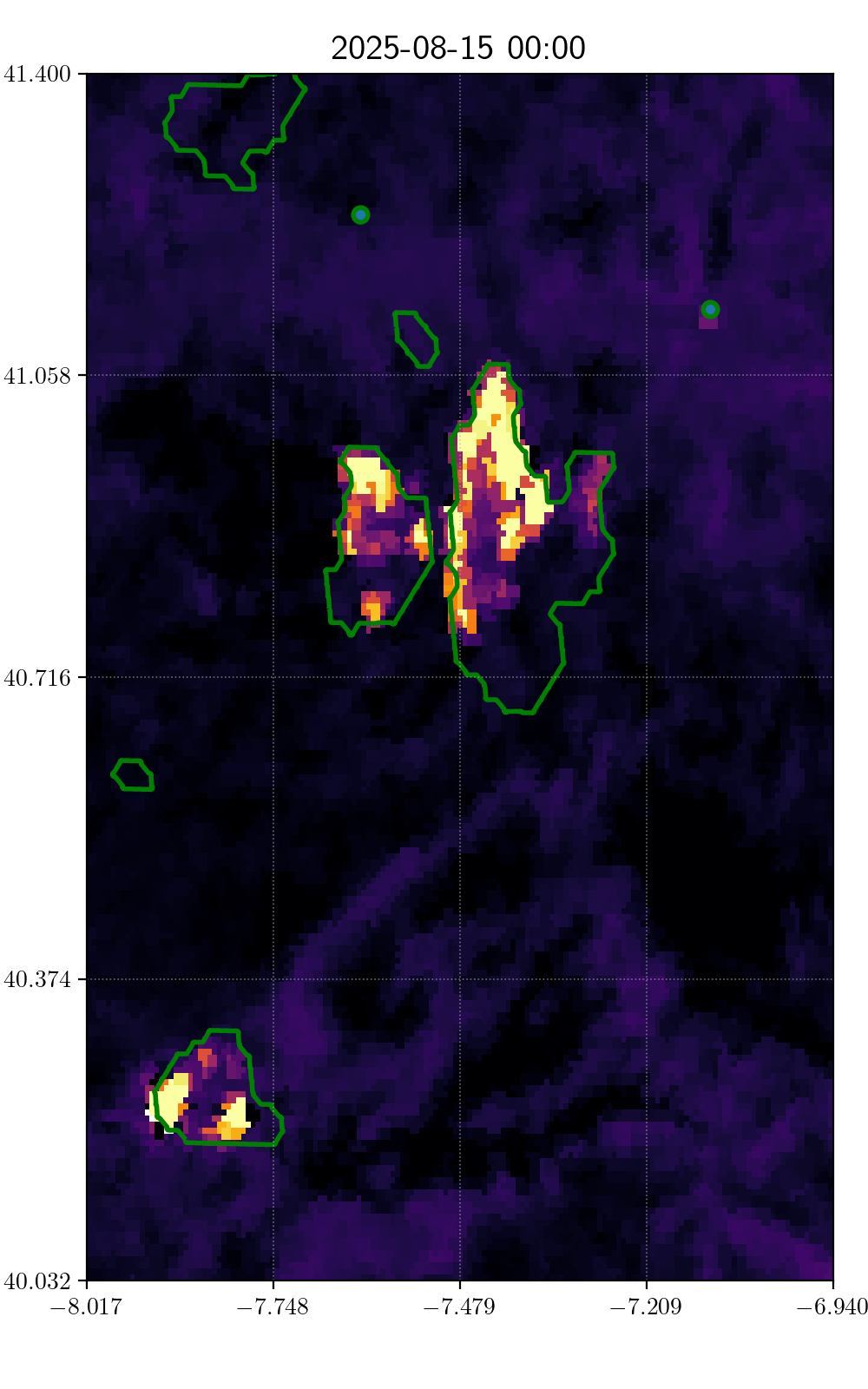}
    \end{subfigure}%
    \begin{subfigure}{0.22\textwidth}
        \includegraphics[width=\linewidth]{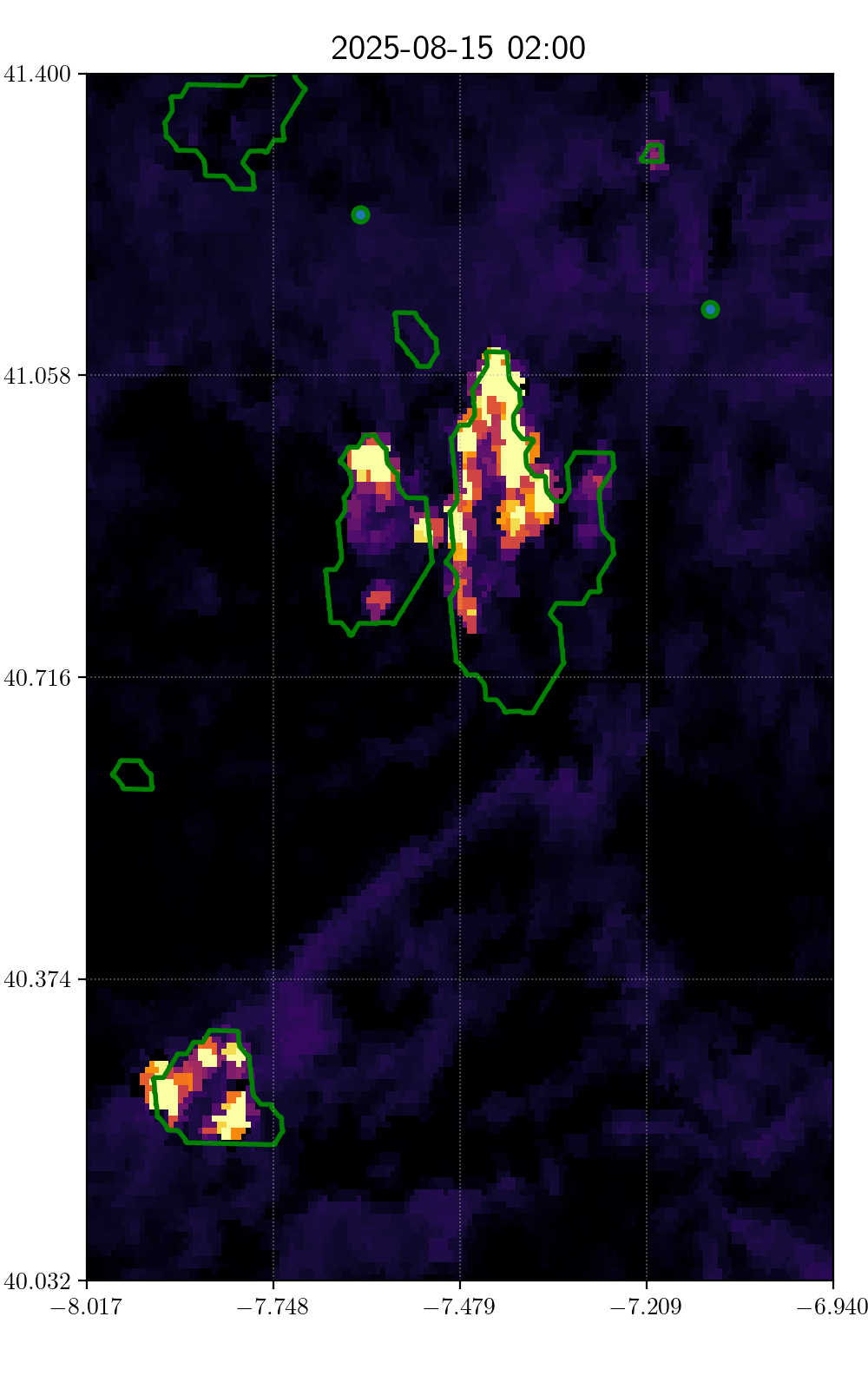}
    \end{subfigure}%
    \begin{subfigure}{0.244\textwidth}
        %\hspace{-2em} % shift image right
        \includegraphics[width=\linewidth]{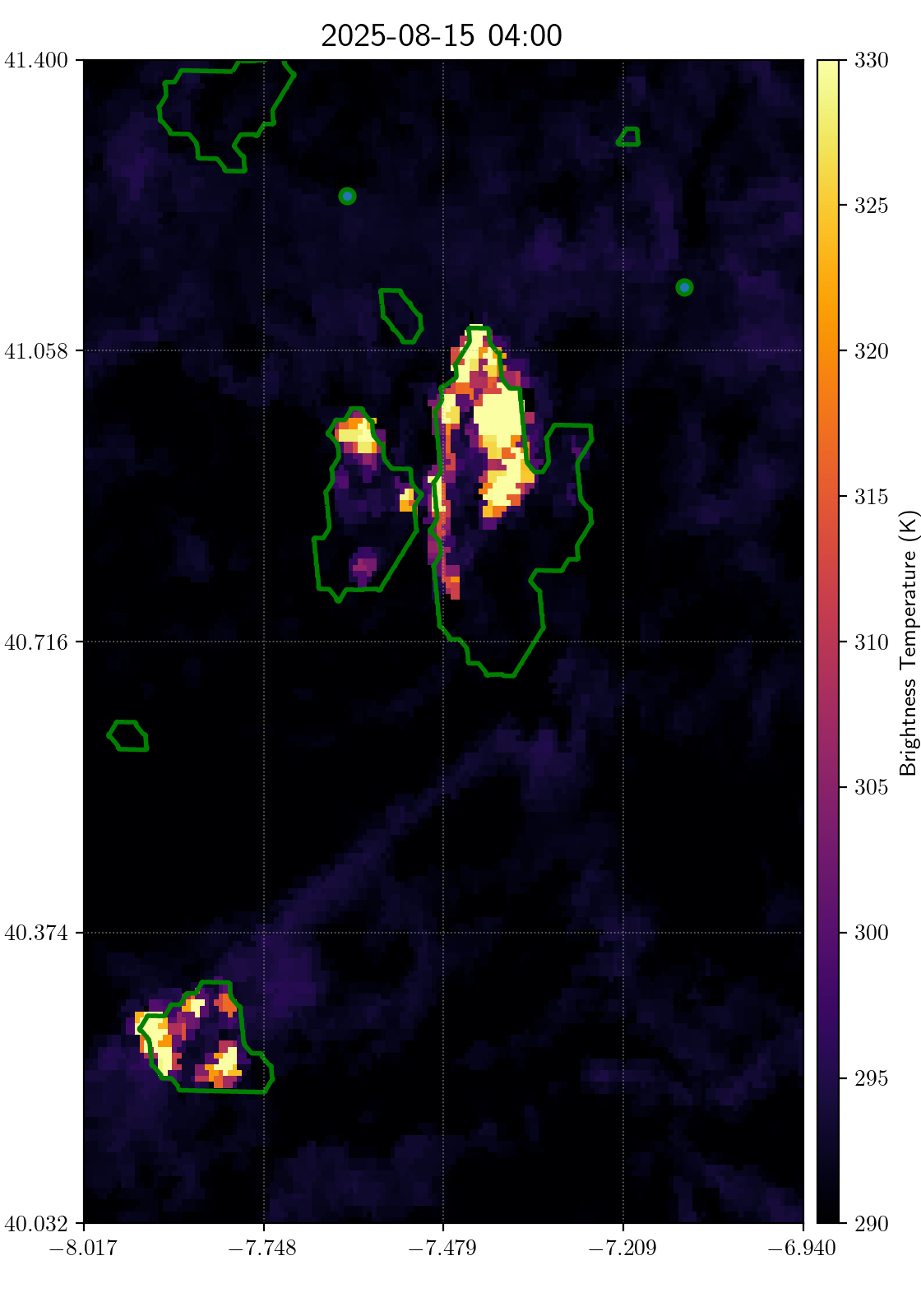}
    \end{subfigure}
    \caption{Update of the Figure \ref{fig:FET_PROTUGAL_PRT} with the alpha shape algorithm implemented in the Fire Event Tracker algorithm. RGB (top) and Middle-Wave Infrared (MWIR, 3.8 µm; bottom) observations of fires in Portugal’s Região Centro, acquired by the FCI sensor onboard Meteosat Third Generation Imager-1 between 14 August 13:00 UTC and 15 August 04:00 UTC. RGB imagery corresponds to daytime hours, while MWIR data are shown for the period after sunset. Yellow and Green outlines (in top and bottom panels respectively) indicate the fire areas of interest as delineated by the Fire-Event-Tracker algorithm.}
    \label{fig:FET_PROTUGAL_MED}
\end{figure}

Among the most recent FET developments that were integrated for the MED domain configuration run, the implementation of the alpha-shape algorithm --which delineates the concave polygon defined by the hotspot cluster-- yielded a clear improvement. Figure \ref{fig:FET_PROTUGAL_MED} presents the updated version of Figure \ref{fig:FET_PROTUGAL_PRT}, covering the same spatial domain and temporal period, but derived from the MED configuration run in which the alpha-shape method was applied. The resulting AOI polygon exhibits substantially greater detail and now more accurately reproduces the geometry of the fire front, as observed in the MWIR and visible-band rasters.

In addition to the potential improvements of the FET algorithm, it is important to note for prospective operational implementation that the data acquisition system covering Portugal has been running since 14 August 2025. It provides measurements every 10 minutes, in synchronization with the sensor acquisition cycle. And as of 14 February 2026, it has achieved an overall fire event data availability of 98\%.

\section{Conclusions}
%-------------------
This study demonstrates the transformative potential of MTG-FCI active fire observations for NRT wildfire intelligence in Europe. By introducing the FET algorithm and implementing it in an operational context during the 2025 fire season in Portugal, as well as during the SILEX airborne campaign, we show that continuous satellite-based fire monitoring can be transitioned from research-focused prototypes to systems that directly support tactical decision-making and contribute to improved operational safety.

The unprecedented temporal resolution and coverage of FCI enable a qualitative shift in fire observation: instead of relying on isolated detections, fire events can now be tracked as evolving systems, with trajectories, intensities, and interactions resolved at scales relevant for suppression agencies and airborne operations. This represents a step change in situational awareness, particularly in densely populated and infrastructure-rich landscapes where timely information are decisive.

Looking forward, the integration of FET into multi-sensor observation chains and coupled fire–atmosphere modeling frameworks opens the prospect of predictive fire intelligence. Such systems could not only monitor ongoing incidents but also anticipate fire spread, plume development, and potential risks to communities and responders. Embedding these capabilities into civil protection infrastructures across Europe would contribute to a more adaptive and resilient approach to fire management in a climate where extreme fire weather will become the norm rather than the exception.

Ultimately, the operational exploitation of MTG-FCI data marks the beginning of a new era in wildfire monitoring. By coupling space-based assets with advanced modeling and decision-support tools, Europe can establish a next-generation fire intelligence system that will offer better monitoring capabilities of the growing wildfire challenge of the 21st century.

\section{Acknowledgements}
%-------------------
This work was partly funded by Grant PID2023-150607OB-I00 funded by MICIU/AEI/ 10.13039/501100011033 and by ERDF/EU. The collaboration with the Autoridade Nacional de Emergência e Proteção Civil of Portugal was supported by an STSM grant (E-COST-GRANT-CA22164-fd0f9b22) of the NERO Cost action CA22164.
The SILEX campaign of the EUBURN project was supported by Météo-France, the French National Research Agency under grant agreement no. ANR-24-CE01-3132, the French national program LEFE/INSU, the French National Agency for Space Studies (CNES) and the European Union under the Interreg Sudoe Programme (S2/2.4/F0327).
We thank AERIS for supporting the development of the SILEX operational web site.
Simulations of FOREFIRE-MESONH were supported by EU H2020 101037419 Fire-Res program and Brando supercomputer center of the University of Corsica.

\section{Supplementary material}
%-------------------
\subsection{Visualization}
\begin{figure}[!ht]
    \centering
    \includegraphics[width=0.5\linewidth]{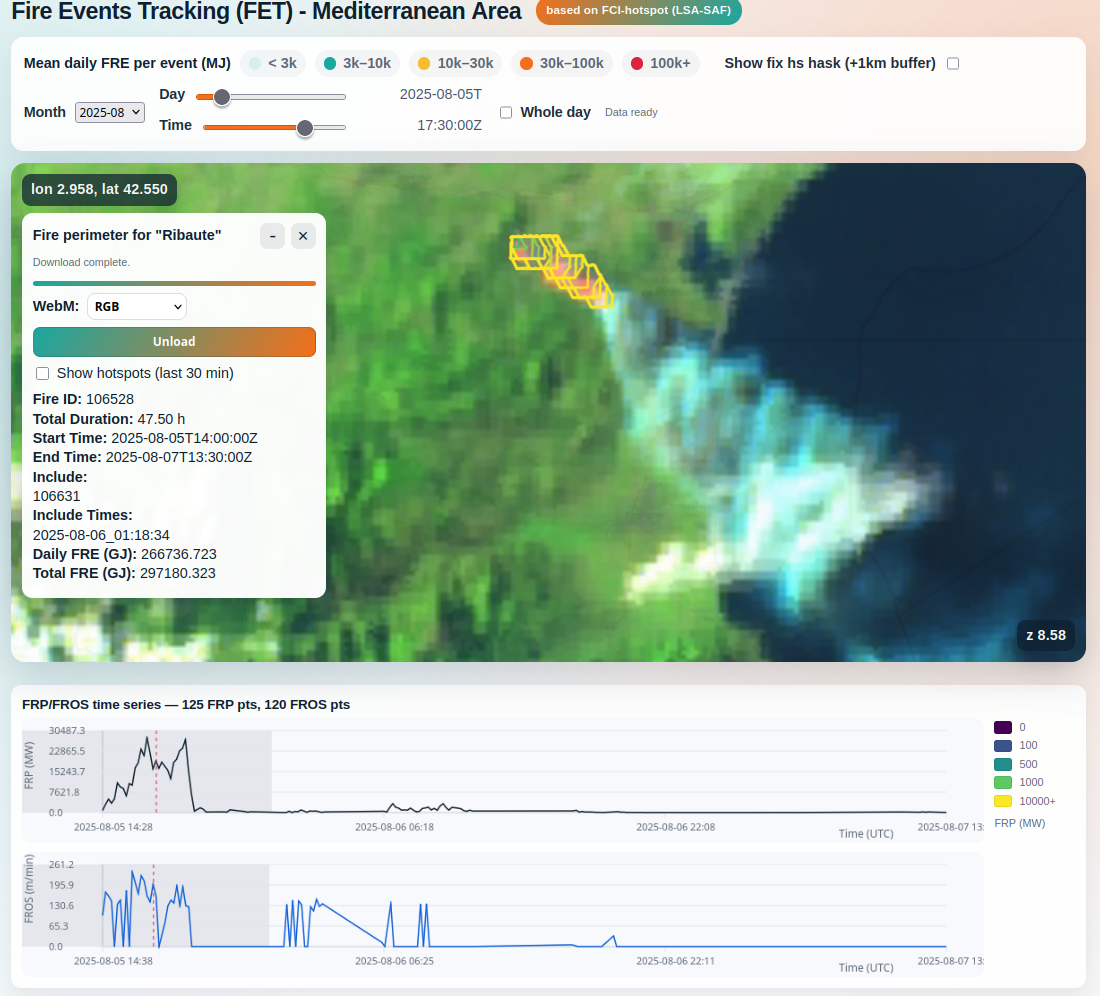}
    \caption{View from the web interface develop to support the Fire Event Tracker algorithm.}
    \label{fig:FETView}
\end{figure}

To facilitate the visualization of the FET data, a web interface was developed to support the retro-analysis of the MED configuration run. Figure \ref{fig:FETView} shows a view of this website that is available at \url{https://certec.mtg.eebe.upc.edu:444/viewer/?time=2025-08-05T14%3A00%3A00Z#4.34/42.38/8.78}.

The FET NRT application for the 2025 fire season 2025 (PRT and SILEX) were also delivered with a web interface.
Only the NRT application of the PRT configuration is still running operationally and is available at \url{htpps://185.32.190.239:8000}. It was developed in collaboration with the association  Associação de Voluntários Digitais em Situações de Emergência (VOST) and used during ANEPC operation over the 2025 fire season.

In the three web interface applications, an effort was made to complement the event tracking with visual context in the visible and Infra red using chunks of the FCI raster data available from the EUMETSAT data store.
The raster are available with a same latency of approximately 20 minutes and delivered with the 10-mon cadence of the sensor.

\subsection{Alpha shape algorithm add on}
To show the improvement generated by the alpha shape algorithm and defining the AOI with a concave polygon instead of the a convex polygon, Figure \ref{fig:FET_EFFIS_MED} present an update of Figure \ref{fig:FET_PROTUGAL_PRT} on the same spatial extent and same times. See the Discussion section for more details.

\section{Appendices}
\subsection{Method for estimating Forward Rate of Spread-FROS}

The rate of spread in each event in FET is computed using the Forward Rate Of Spread (FROS) approach of \cite{Benali2023}.
At each time step, for each event, it estimates propagation vector between the last two successive polygonal perimeters by analyzing their spatial nesting and geometric separation over time.
To improve the spatial sampling of polygon perimeter, each polygon/line is densified by inserting intermediate vertices along each edge. This step increases the number of candidate points used in distance calculations and reduces bias associated with coarse vertex spacing.
For each pair of last two successive polygons, the algorithm evaluates whether one polygon (denoted as the ``outer'' polygon at time $t_1$) fully contains another polygon (the ``inner'' polygon at time $t_2$, with $t_1 > t_2$). Only such nested pairs are considered, ensuring that the analysis reflects outward propagation.

To estimate spatial propagation, the method computes distances between boundary vertices of the outer and inner polygons.
All candidate point pairs are connected by line segments, and each segment is subjected to a set of geometric constraints:
\begin{enumerate}
    \item The segment must lie entirely within the outer polygon.
    \item The segment must not be fully contained within the inner polygon.
    \item The segment must not intersect the boundary of any other polygon (treated as obstacles), except at its endpoints within a specified tolerance.
\end{enumerate}
Segments violating any of these conditions are discarded. For the remaining valid segments, Euclidean distances between the corresponding vertex pairs are computed. Extremely large distances relative to the time interval are filtered out to avoid spurious estimates.

In addition to the scalar velocity, a representative propagation direction is derived. This is achieved by constructing line segments between corresponding vertex pairs and computing a mean vector line based on the average start and end coordinates.

This vector approach provides a fast and geometrically constrained estimate of propagation that accounts for complex polygon shapes and excludes unrealistic paths caused by intersections or topological inconsistencies.

\subsection{FROS performance analysis}

\begin{figure}[!ht]
    \centering
    \includegraphics[width=.5\linewidth]{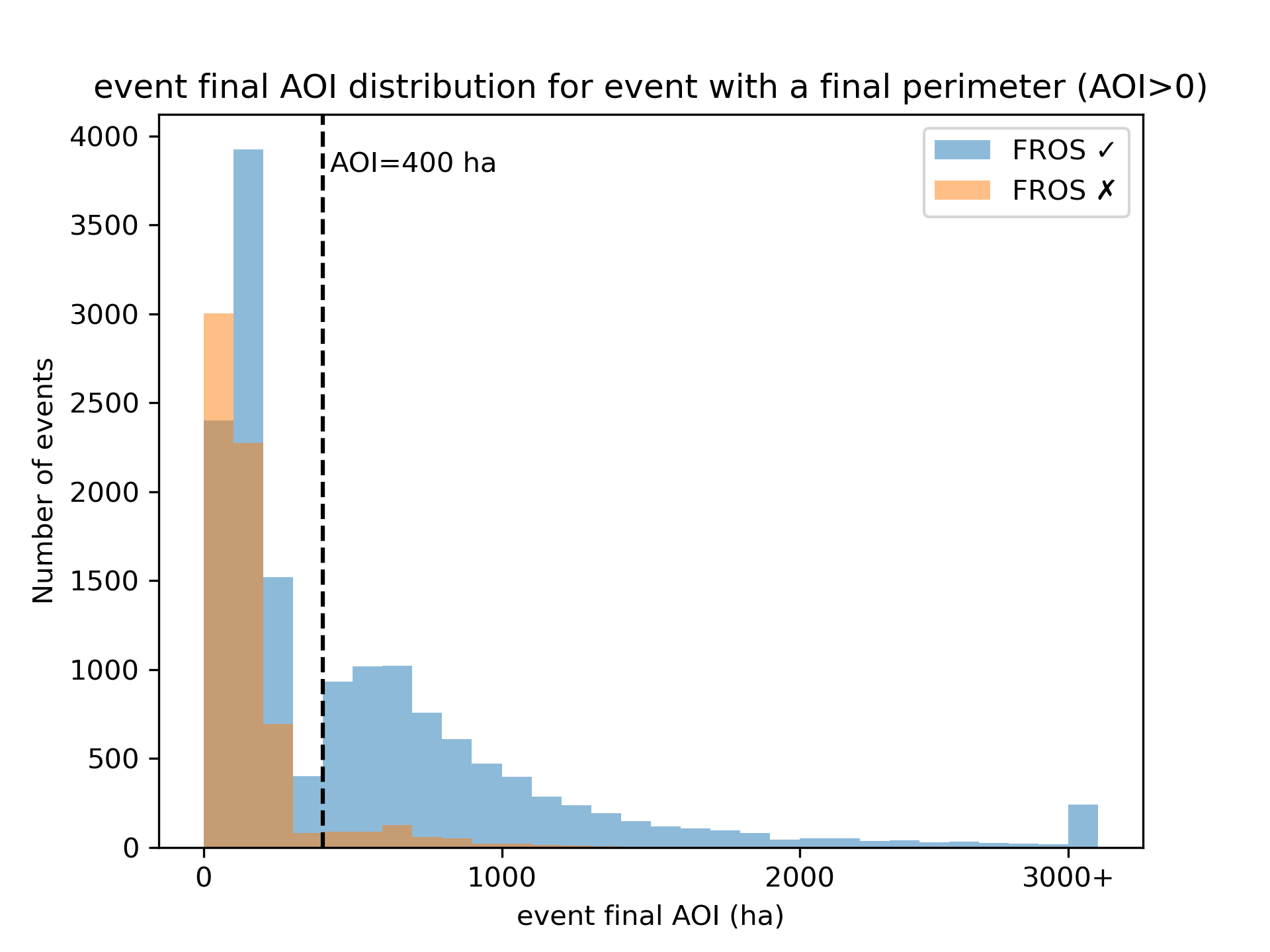}
    \caption{Distribution of the Area Of Interest (AOI) for fire events in which the FROS computation is successful (blue) and in which it is unsuccessful (orange).}
    \label{fig:MED3_AOIDistribution}
\end{figure}

This section presents an analysis of the performance of the FROS algorithm with the objective of identifying a critical AOI size beyond which the FET algorithm becomes more reliable. Figure \ref{fig:MED3_AOIDistribution} shows the distribution of AOI for fire events in which the FROS algorithm operated successfully and those in which it failed. For fires larger than $400~\text{ha}$, there is a clear increase in the success rate of FROS calculations. This improvement is associated with a more coherent and stable organization of the fire perimeter as fire size increases, which in turn better satisfies the geometric constraints imposed by the algorithm. Consequently, $400~\text{ha}$ is adopted as a lower threshold beyond which FCI exhibits enhanced capability to correctly capture the geometry of the fire front.

\bibliographystyle{plain}
\bibliography{library,library_ronan}

\end{document}